\documentclass[a4paper,12pt]{article}            
\usepackage{jheppub}
\pdfoutput=1

\usepackage[utf8]{inputenc}
\usepackage{amsmath,amssymb}
\usepackage[normalem]{ulem}
\usepackage{xcolor}
\usepackage{tensor}
\usepackage{graphicx}
\usepackage{subcaption}
\usepackage{xurl} 
\usepackage{hyperref}
\usepackage{bbm}

\title{When duality changes the poles: $SL(2,\mathbb{Z})$ transformations of linear response EFTs}
\author[1,2]{Andrea Amoretti,}
\author[1,2]{Daniel K. Brattan,} 
\author[1,2]{Jonas Rongen}

\emailAdd{andrea.amoretti@ge.infn.it}
\emailAdd{danny.brattan@gmail.com}
\emailAdd{jonas.ludovico.rongen@edu.unige.it}

\affiliation[1]{Dipartimento di Fisica, Universit\`a di Genova,
via Dodecaneso 33, I-16146, Genova, Italy}
\affiliation[2]{I.N.F.N. - Sezione di Genova, via Dodecaneso 33, I-16146, Genova, Italy}

\begin{abstract}
{\ We study how $SL(2,\mathbb Z)$ transformations act on pole-truncated linear-response effective theories in $(2+1)$ dimensions. For correlators meromorphic within a finite low-frequency domain, the EFT data include both derivative-expanded transport coefficients and the retained non-hydrodynamic poles. Since the $S$-trans\-for\-mation inverts the optical conductivity and exchanges its poles and zeros, it does not necessarily commute with the infrared truncation. We derive explicit maps for static susceptibilities, the Einstein relation and AC transport coefficients. In the purely hydrodynamic sector the map closes on the original transport data, while in quasihydrodynamic regimes the transformed pole positions and, in general, residues constitute additional infrared data. We test the construction in the alternatively quantised D3/D5 probe-brane system, where an EFT retaining one original gapped pole can correspond to transformed EFTs retaining zero, one or two. The holographic results verify the analytic maps and exhibit an inverse-Drude conductivity at large values of the control parameter.}
\end{abstract}
\begin{document}

\maketitle

\section{Introduction}

Dualities and effective field theories are two of the most powerful organising
principles in theoretical physics, but they encode information in very different
ways. A duality is an exact statement about a microscopic theory, or equivalently
about its complete set of correlation functions. An effective field theory, by
contrast, deliberately discards most of that information and keeps only what is
needed to describe the infrared. It is therefore natural to ask whether these two
structures are compatible: does an exact duality of the microscopic theory descend
to a simple transformation of its low-energy effective description? In this paper we
show that, for quasihydrodynamic effective theories in $(2+1)$ dimensions, the answer
is generically no, and we identify precisely why.

Effective field theories of hydrodynamics are usually organised as derivative expansions around conserved densities~\cite{Dubovsky:2011sj,Crossley:2015evo,Glorioso:2016gsa,Kovtun:2014hpa,Davison:2014lua,Davison:2015bea,Donos:2018kkm}. In strongly coupled systems, however, long-lived non-hydrodynamic excitations may also have to be retained explicitly~\cite{Grozdanov:2018fic,Baggioli:2023mid,Liu:2024tqe,Heller:2020uuy,Brattan:2024dfv}. The framework developed in \cite{Amoretti:2025kem} describes such linear response through a Mittag-Leffler expansion, so that the low-energy data consist not only of derivative-expanded transport coefficients but also of the positions and residues of the poles retained inside a chosen frequency domain. These pole data are non-perturbative from the low-frequency perspective, while the extent of the low-energy domain is constrained by the analytic structure of the correlator \cite{Withers_2018,Grozdanov:2019kge,Grozdanov:2019uhi}.  More general non-hydrodynamic analytic structures, including branch cuts obtained from continua of relaxation modes, can also be incorporated within Schwinger-Keldysh effective theory~\cite{Amoretti:2026branch}, but lie outside the finite meromorphic truncations considered here.

An exact transformation acting on the complete correlator does not, in general, preserve this finite pole content. The relevant question is therefore how to construct the transformed EFT when the set of poles retained inside the low-frequency domain changes. Our main result is an explicit dictionary for this construction. In the purely hydrodynamic sector the map closes on the transport coefficients of the original EFT. When long-lived non-hydrodynamic modes are retained, the positions and, in general, residues of the transformed poles must additionally be supplied. These constitute non-perturbative infrared data and cannot in general be reconstructed from a finite low-frequency expansion of the original EFT.

We study this issue using $SL(2,\mathbb Z)$ transformations of $(2+1)$-dimensional quantum field theories~\cite{Witten:2003ya,Leigh:2003ez}. These transformations mix the electric current with the magnetic current and are closely related to particle-vortex duality~\cite{Turner:2019wnh,Metlitski:2015eka}. In particular, at vanishing momentum in a parity-invariant theory the $S$-transformation, one of the generators of $SL(2,\mathbb{Z})$, inverts the optical conductivity~\cite{Herzog:2007ij,Sachdev:2011wg}. Holographically it is implemented by changing the boundary condition of the bulk gauge field, or equivalently by passing from standard to alternative quantisation~\cite{Witten:2003ya,Leigh:2003ez,Hartman:2008dq}. For the D3/D5 system considered below this relates two distinct boundary theories and should not be interpreted as a self-duality of the probe-brane action.

Since the $S$-transformation exchanges poles and zeros of the conductivity, applying it to the complete response function and truncating to a fixed low-frequency domain does not necessarily commute. We derive the corresponding maps for static susceptibilities, the Einstein relation and low-frequency AC transport coefficients, and illustrate the resulting pole-counting regimes in the alternatively quantised D3/D5 probe-brane system. The holographic calculation provides explicit checks of the analytic construction and, at large values for the transformed magnetic field, yields the inverse-Drude behaviour expected from the transformed correlator.

The paper is organised as follows. In Section~\ref{section:S-Field} we review the
$SL(2,\mathbb Z)$ action on $(2+1)$-dimensional field theories, keeping only the
ingredients needed for linear response. In Section~\ref{sec:eft} we recall the
pole-data definition of the linearised effective theory and construct the
$S$-transformed effective description; in particular, in
Subsection~\ref{sec:Stransformation} we derive the maps for susceptibilities,
diffusion and AC transport coefficients. In Section~\ref{sec:D3D5} we apply the
construction to the alternatively quantised D3/D5 probe brane system and show
explicitly the one-to-one, one-to-two and one-to-zero pole regimes. We conclude in
Section~\ref{sec:discussion} with the conceptual implications of our results and
possible extensions.

\section{$SL(2,\mathbb{Z})$ transformation of field theory}
\label{section:S-Field}

\noindent
Let us begin by recalling the action of $SL(2,\mathbb Z)$ transformations on
a $(2+1)$-dimensional quantum field theory with a conserved, gauge-invariant
$U(1)$ current $J^\mu$ coupled to a background gauge field $A_\mu$.
Following \cite{Witten:2003ya,Leigh:2003ez}, these transformations are generated
by two elementary operations. The $S$-transformation promotes the background gauge
field to a dynamical degree of freedom and simultaneously introduces a new background gauge field $A_\mu^* $, while the $T$-transformation adds a Chern-Simons contact term for the background field.
Together, $S$- and $T$-transformations generate the full $SL(2,\mathbb Z)$ group.

Since our interest is in linear response, we will only need the induced action on
currents, background fields and two-point functions. We refer to
Appendix~\ref{app:sl2z} for the derivation from the generating functional.

In the following, we use the convention $\epsilon_{txy}=1$, and define the magnetic current
associated with a background gauge field $A_\mu$ by
\begin{equation}
    \mathcal B^\mu[A]
    =
    \frac{1}{2\pi}\epsilon^{\mu\nu\rho}\partial_\nu A_\rho .
\end{equation}
In components this becomes
\begin{equation}
    \mathcal B^\mu[A]
    =
    \frac{1}{2\pi}
    \left(
        -B,\,
        \epsilon^{ij}E_j
    \right) , 
\end{equation}
where $B$ and $E_i$ are the magnetic and electric fields constructed from
$A_\mu$. Analogously, in the transformed theory we define
\begin{equation}
    \mathcal B^{\mu *}
    \equiv
    \mathcal B^\mu[A^*]
    =
    \frac{1}{2\pi}\epsilon^{\mu\nu\rho}\partial_\nu A^*_\rho . 
\end{equation}

We first focus on the pure $S$-transformation, under which the original background field $A_\mu$ becomes dynamical. The transformed theory is then sourced by the new background field $A_\mu^*$. The current of the new theory is the magnetic current of the original gauge field,
\begin{equation}
    J^{\mu *} = \mathcal B^\mu[A] . 
\end{equation}
The equation of motion of the dynamical field $A_\mu$ relates the magnetic
current of the transformed theory to the original electric current,
\begin{equation}
    \mathcal B^{\mu *} = - J^\mu .
\end{equation}
More compactly, we can write 
\begin{equation}
    \begin{pmatrix}
        J^{\mu *} \\
        \mathcal B^{\mu *}
    \end{pmatrix}
    =
    \begin{pmatrix}
        0 & 1 \\
        -1 & 0
    \end{pmatrix}
    \begin{pmatrix}
        J^\mu \\
        \mathcal B^\mu
    \end{pmatrix} , 
\end{equation}
which becomes in components 
\begin{equation}
    \rho^* = -\frac{B}{2\pi},
    \qquad
    J^{i*} = \frac{1}{2\pi}\epsilon^{ij}E_j,
    \qquad
    B^* = 2\pi \rho,
    \qquad
    E^{i*} = 2\pi \epsilon^{ij}J_j , 
\end{equation}
where $\rho\equiv J^t$ and $\rho^*\equiv J^{t*}$. Thus the $S$-transformation exchanges electric and magnetic degrees of freedom.

The $T$-transformation instead leaves the background field non-dynamical and
adds a Chern-Simons contact term. At the level of currents this shifts the electric
current by an integer multiple of the magnetic current,
\begin{equation}
    J^{\mu *}=J^\mu+n\,\mathcal B^\mu,
    \qquad
    \mathcal B^{\mu *}=\mathcal B^\mu,
    \qquad
    n\in\mathbb Z .
\end{equation}
Since $S$ and $T$ do not commute, as mentioned above, their products generate a general
$SL(2,\mathbb Z)$ transformation. For later use, we write its action as
\begin{subequations}
\label{eq:SL-Transformation}
    \begin{eqnarray}
        \begin{pmatrix}
            J^{\mu *} \\
            \mathcal B^{\mu *}
        \end{pmatrix}
        =
        \begin{pmatrix}
            a & b \\
            c & d
        \end{pmatrix}
        \begin{pmatrix}
            J^\mu \\
            \mathcal B^\mu
        \end{pmatrix} , 
    \end{eqnarray}
    with
    \begin{eqnarray}
        ad-bc=1,
        \qquad
        a,b,c,d\in\mathbb Z . 
    \end{eqnarray}
\end{subequations}
The pure $S$-transformation corresponds to
\begin{equation}
    a=d=0,
    \qquad
    b=-c=1 .
\end{equation}

Using the component form of the magnetic current, the general transformation
\eqref{eq:SL-Transformation} gives
\begin{equation}
    \begin{aligned}
    \label{eq:SL-Field-Transformations}
        \rho^*
        &=
        a\rho-\frac{b}{2\pi}B,
        &
        J^{i*}
        &=
        aJ^i+\frac{b}{2\pi}\epsilon^{ij}E_j,
        \\
        B^*
        &=
        -2\pi c\,\rho+dB,
        &
        E^{i*}
        &=
        dE^i-2\pi c\,\epsilon^{ij}J_j,
    \end{aligned}
\end{equation}
with the inverse relations being
\begin{equation}
    \begin{aligned}
    \label{eq:Field-SL-Transformations}
        \rho
        &=
        d\rho^*+\frac{b}{2\pi}B^*,
        &
        J^i
        &=
        dJ^{i*}-\frac{b}{2\pi}\epsilon^{ij}E_j^*,
        \\
        B
        &=
        aB^*+2\pi c\,\rho^*,
        &
        E^i
        &=
        aE^{i*}+2\pi c\,\epsilon^{ij}J_j^* .
    \end{aligned}
\end{equation}

These transformations induce corresponding relations between current-current
correlators. In particular, in parity-invariant theories the pure
$S$-transformation exchanges the longitudinal and transverse sectors and maps
each response function to the inverse of the other, up to a conventional factor
of $2\pi$. At vanishing spatial momentum this reduces to the inversion
of the optical conductivity,
\begin{equation} \label{eq:S_conductivity}
    \sigma^*(\omega)
    =
    \frac{1}{(2\pi)^2}\frac{1}{\sigma(\omega)} .
\end{equation}
Therefore, poles of the conductivity of the original theory are mapped to zeros of
the transformed conductivity, and zeros are mapped to poles.

Because a low-energy EFT retains only the poles inside a chosen frequency domain, this inversion does not, in general, preserve the pole content of a pole-truncated description. We focus below on the pure $S$-transformation relevant to the D3/D5 application and return briefly to general $SL(2,\mathbb Z)$ transformations in Section~\ref{Sec:SL2Zgeneral}.

\section{Effective linearised theory}
\label{sec:eft}

\begin{figure}[t]
\centering
\includegraphics[width=0.45\textwidth]{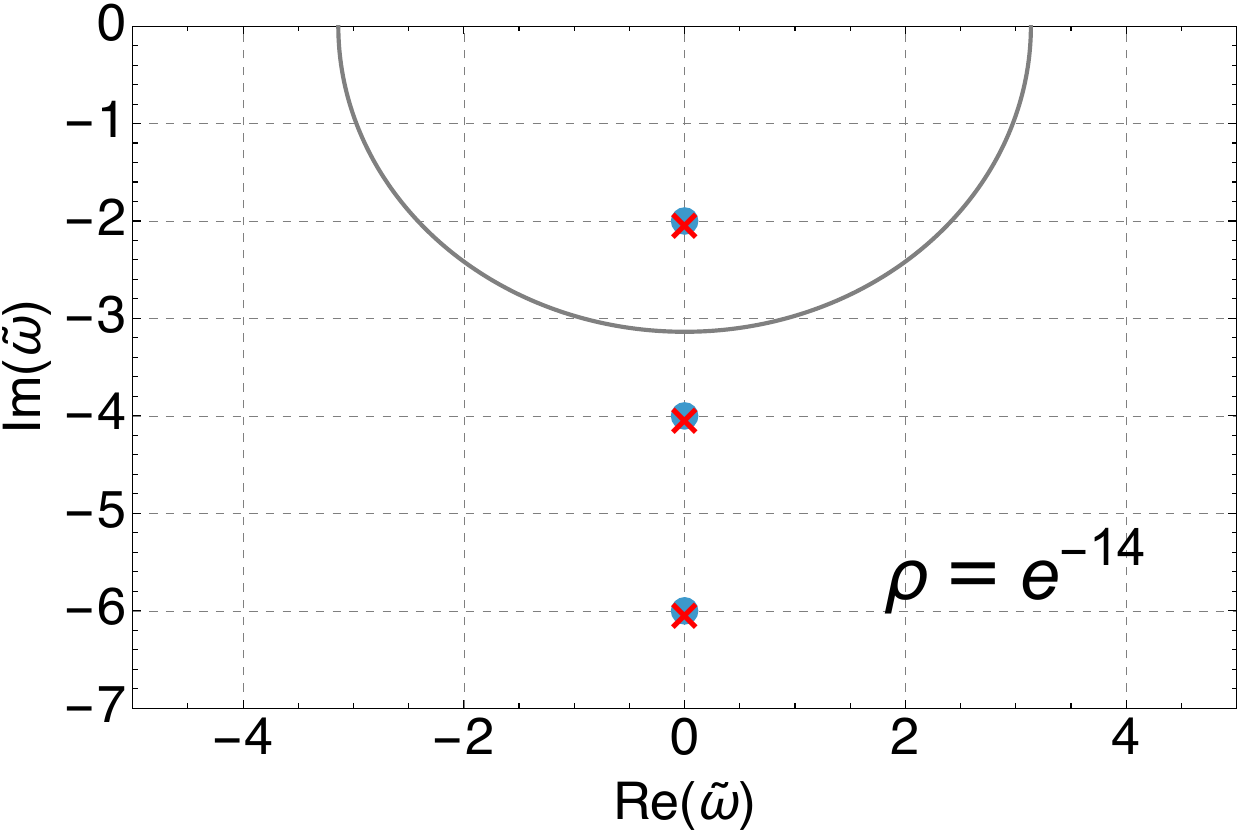}
\includegraphics[width=0.45\textwidth]{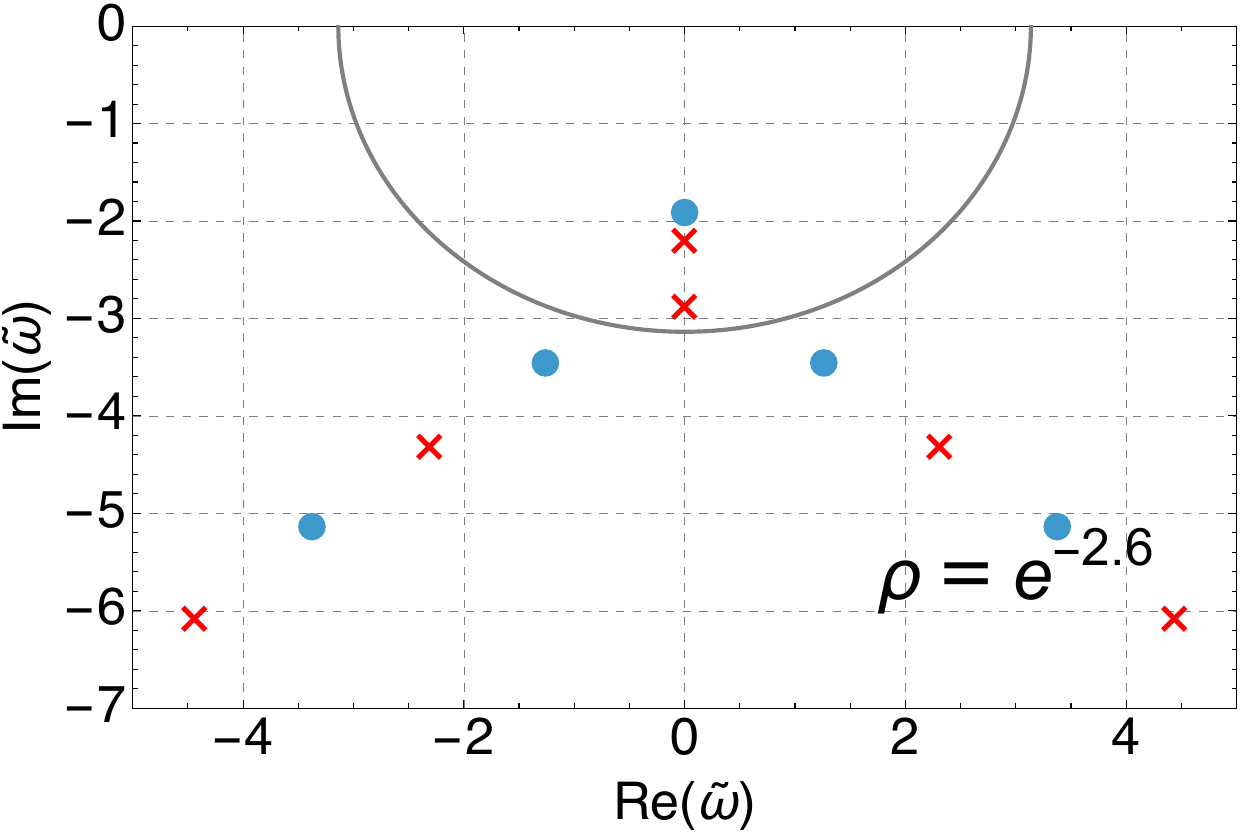}
\includegraphics[width=0.45\textwidth]{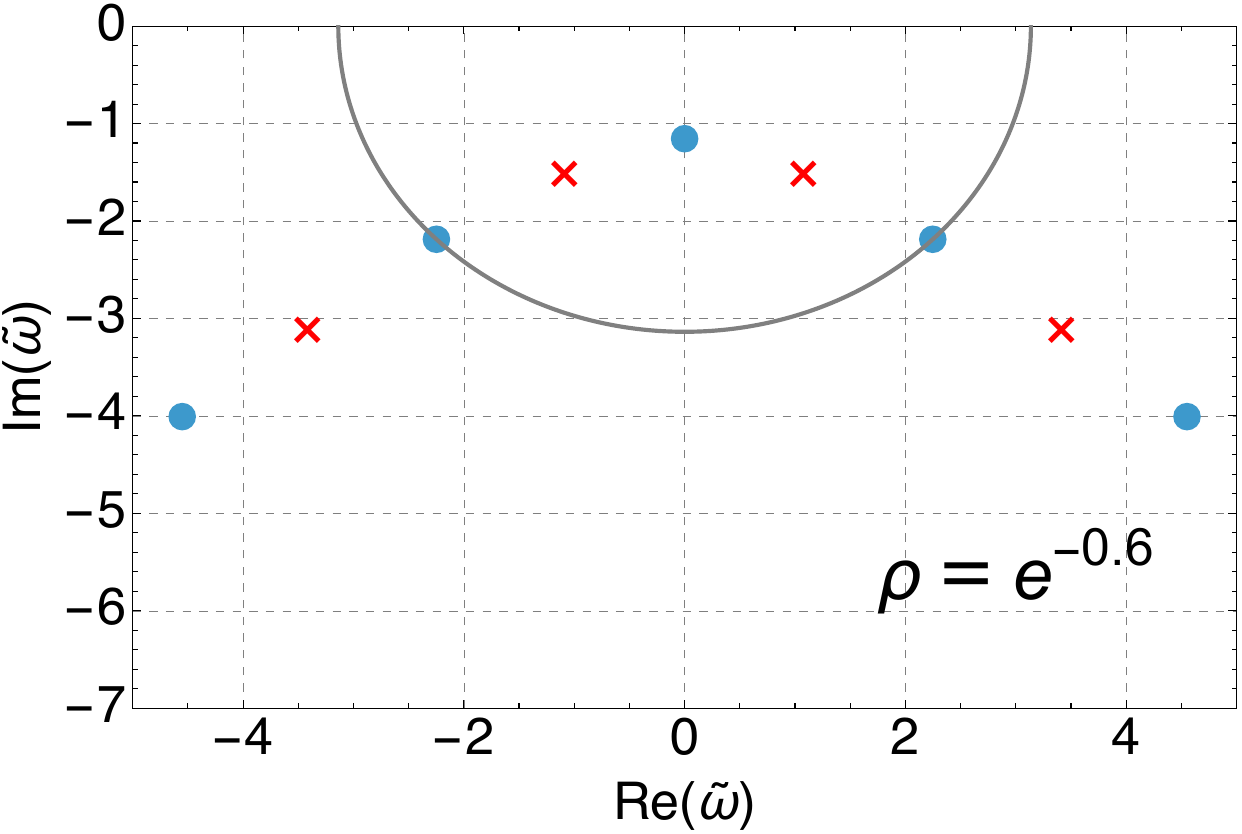}
\includegraphics[width=0.45\textwidth]{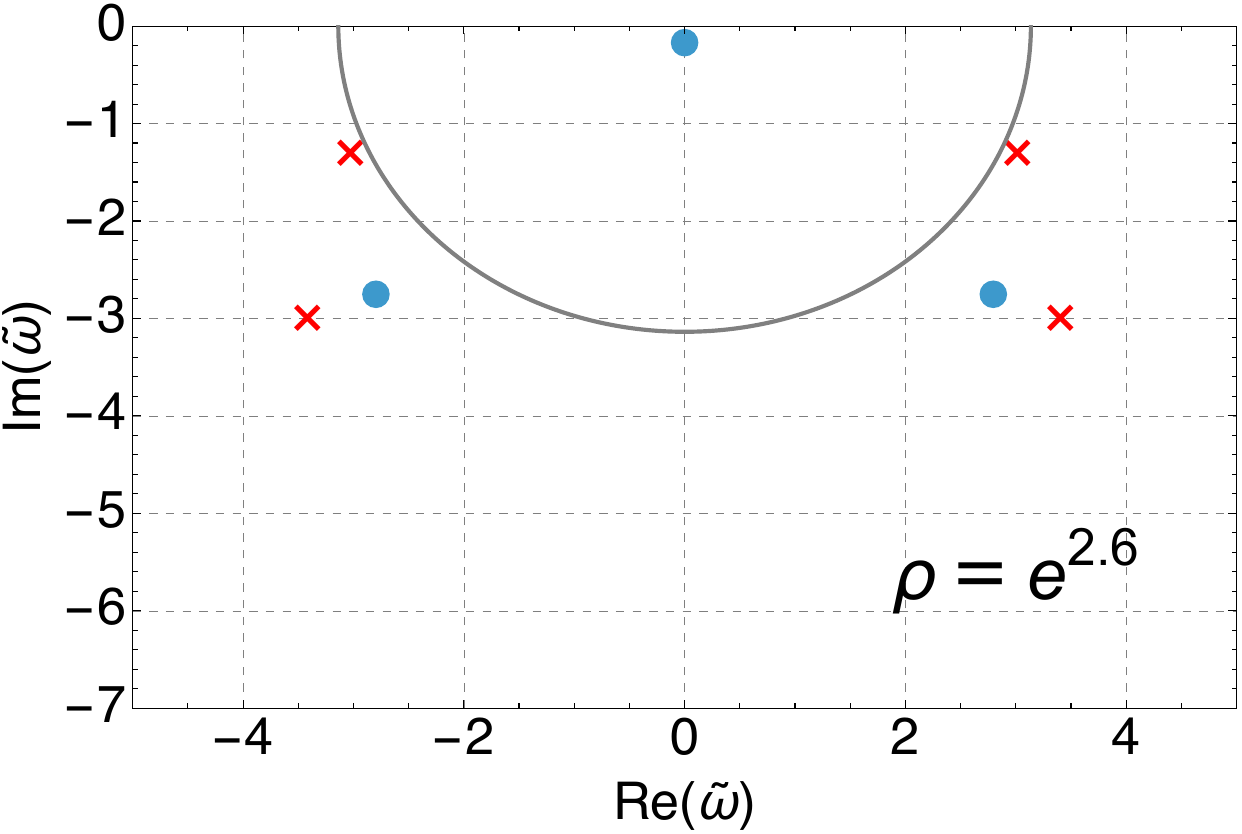}
\caption{The positions of the gapped poles of the D3/D5 current-current
correlator at vanishing wavevector in the original theory, shown as blue dots,
and in the $S$-transformed theory, shown as red crosses, for different charge
densities. The grey semi-circle indicates the frequency domain used in the
effective description.}
\label{fig:poles}
\end{figure}

The effective framework of \cite{Amoretti:2025kem} treats the gapped poles retained inside a chosen low-frequency domain as part of the EFT input, alongside the derivative-expanded transport coefficients. Figure~\ref{fig:poles} illustrates that the number of such poles inside the same domain, in general, does not agree before and after the $S$-transformation. For the representative charge densities shown in this figure, the transformed theory can contain a different number of gapped poles inside the same effective domain. Thus, transforming an already pole-truncated EFT usually does not reproduce the EFT obtained by first transforming the full correlator.

We now recall the linearised effective theory. We assume that the retarded correlators are meromorphic inside the chosen domain, which may contain any finite number of isolated simple gapped poles, so that they admit a Mittag-Leffler representation consisting of the retained pole contributions plus a holomorphic part. The constitutive relation for the charge current is
\begin{equation}
\label{eq:constitutive1}
    \delta J^\mu
    =
    \left(
        \chi_{\rho\rho}\delta\mu
        -
        \nabla^\perp_\nu \delta p^\nu
    \right) u^\mu
    +
    \left(
        \delta\bar J^\mu
        -
        {\Pi^\mu}_\nu \nabla_\rho \delta M^{\rho\nu}
    \right) ,
\end{equation}
where $\delta M^{\mu \nu}$ is defined as 
\begin{equation}
    \delta M^{\mu\nu}
    =
    u^\nu\delta p^\mu
    -
    u^\mu\delta p^\nu
    -
    \epsilon^{\mu\nu\rho}u_\rho\delta m .
\end{equation}
Here $\delta p^\mu$ and $\delta m$ are the polarisation and magnetisation
fluctuations, with $\delta p^\mu=u_\rho \delta M^{\rho\mu}$. The tensor
${\Pi^\mu}_\nu$ denotes the projector orthogonal to the fluid velocity such that 
\begin{equation}
    \nabla^\perp_\mu={\Pi_\mu}^\nu\nabla_\nu . 
\end{equation}
Furthermore we use the standard normalisation condition $ u^\mu u_\mu = - 1$. 
We work in a frame in which derivative corrections to $\delta J^\mu$ are
entirely transverse to the fluid velocity at all orders in derivatives. Further details of the construction
can be found in \cite{Amoretti:2025kem}.

The equations of motion are
\begin{subequations}
\label{eq:eom}
\begin{align}
    \chi_{\rho\rho}D\delta\mu
    +
    \nabla^\perp_\mu\delta\bar J^\mu
    &=0 ,
    \label{eq:constitutive2a}
    \\
    \left[
        \prod_{n=1}^{N-1}
        \left(
            \Pi^{\mu\nu}D+\Gamma_n^{\mu\nu}[\nabla_\perp^2]
        \right)
    \right]\delta\bar J_\nu
    &=
    \bar\sigma^{\mu\nu}[\nabla_\perp^2]
    \left(
        \delta E_\nu-\nabla^\perp_\nu\delta\mu
    \right)
    \nonumber
    \\
    &\quad
    +
    r^{\mu\nu}[D,\nabla_\perp^2]
    \left(D\delta E_\nu\right) ,
    \label{eq:constitutive2b}
\end{align}
\end{subequations}
where the electric and magnetic fields are
\begin{subequations}
\begin{align}
    \delta B
    &=
    -\frac12 u_\mu\epsilon^{\mu\nu\rho}
    \left(
        \partial_\nu\delta a_\rho
        -
        \partial_\rho\delta a_\nu
    \right),
    \\
    \delta E_\mu
    &=
    u^\nu
    \left(
        \partial_\mu\delta a_\nu
        -
        \partial_\nu\delta a_\mu
    \right).
\end{align}
\end{subequations}
The first equation in \eqref{eq:eom} is charge conservation. In the second
equation, the spatial part of the current has been promoted to an independent
variable $\delta\bar J^\mu$ obeying a relaxation equation. This is analogous
to Müller-Israel-Stewart theory, where the viscous stress tensor is promoted to
an independent variable and produces additional non-hydrodynamic poles. The
operator $D=u^\mu\nabla_\mu$ will later play the role of the time derivative.

In hydrostatic configurations the right-hand side of
\eqref{eq:constitutive2b} vanishes, implying $\delta\bar J^\mu=0$. The
constitutive relations then reduce to those derived from a hydrostatic
generating functional, making the formalism compatible with hydrostaticity. Away from hydrostaticity, the product operator on the left-hand side of \eqref{eq:constitutive2b} 
introduces $N-1$ gapped poles, while charge conservation supplies the diffusive pole in the longitudinal
sector. Quasihydrodynamics is obtained when one of the gapped poles is
parametrically closer to the origin than all other non-hydrodynamic modes.

In the following we choose
\begin{equation}
    u^\mu=(1,\vec 0),
    \qquad
    \vec k=(k,0),
\end{equation}
and use the Fourier convention
\begin{equation}
    \partial_t\to -i\omega,
    \qquad
    \partial_x\to ik .
\end{equation}
Then the linearised field fluctuations become
\begin{subequations}
\label{eq:FluctuationsComponents}
\begin{align}
    \delta B
    &=
    -ik\,\delta a_y ,
    \\
    \delta E_x
    &=
    i(k\delta a_t+\omega\delta a_x) ,
    \\
    \delta E_y
    &=
    i\omega\delta a_y .
\end{align}
\end{subequations}

\subsection{\texorpdfstring{$S$-transformed effective theory}{S-transformed effective theory}}
\label{sec:Stransformation}

We now use this formalism to construct the effective description of the $S$-transformed theory. As should be clear by now, the key point is that this transformed EFT is not obtained, in general, by simply applying the transformation rules of
Section~\ref{section:S-Field} to the finite set of parameters of the original EFT. Rather, the correct procedure is:
\begin{enumerate}
    \item choose a frequency domain $D_R$;
    \item determine the poles of the original microscopic correlator inside
    $D_R$;
    \item apply the exact $S$-transformation to the microscopic correlator;
    \item determine the poles of the transformed correlator inside $D_R$;
    \item build the transformed EFT with this transformed pole data.
\end{enumerate} 
These operations do not need to commute with pole truncation. 

Note that the effective formalism that we reviewed above also applies to the transformed theory, with transformed currents, sources, and transport coefficients and, in general, a different number of retained gapped poles. We denote transformed quantities by an asterisk and focus first on the pure $S$-transformation, returning to general $SL(2,\mathbb Z)$ transformations in Section \ref{Sec:SL2Zgeneral}.

\subsubsection{Susceptibilities}
\label{sec:suscmapping}

We start from the static, finite-momentum constitutive relations of the
original theory,
\begin{subequations}
\label{eq:constitutive_original}
\begin{align}
    \delta J^t
    &=
    \left(
        \chi_{\rho\rho}
        +
        k^2\chi_{\text{EE}}^{(0)}
        +
        k^4\chi_{\text{EE}}^{(\text{L})}(k^2)
    \right)\delta a_t ,
    \\
    \delta J^y
    &=
    k^2\chi_{\text{BB}}(k^2)\delta a_y .
\end{align}
\end{subequations}
Here $\chi_{\rho\rho}$ is the charge susceptibility,
$\chi_{\text{EE}}^{(0)}$ is the momentum-independent part of the electric
polarisation susceptibility, and $\chi_{\text{EE}}^{(\text{L})}$ denotes its
longitudinal momentum-dependent part. The magnetisation susceptibility is
defined by
\begin{equation}
    \delta m=\chi_{\text{BB}}\delta B .
\end{equation}
Equivalently, the retarded Green's functions are
\begin{subequations}
\label{eq:Greens_original}
\begin{align}
    \langle J^t J^t\rangle_{\mathrm R}(0,\vec k)
    &=
    -
    \left(
        \chi_{\rho\rho}
        +
        k^2\chi_{\text{EE}}^{(0)}
        +
        k^4\chi_{\text{EE}}^{(\text{L})}(k^2)
    \right),
    \\
    \langle J^y J^y\rangle_{\mathrm R}(0,\vec k)
    &=
    k^2\chi_{\text{BB}}(k^2).
\end{align}
\end{subequations}
These susceptibilities are obtained by expanding at small $k^2$. Since this
static sector does not involve the gapped relaxation equations, the number of
retained gapped poles does not affect the susceptibility maps.

Using \eqref{eq:FluctuationsComponents} at $\omega=0$, and then the pure
$S$-transformation relations from Section~\ref{section:S-Field}, the original
constitutive relations can be rewritten as
\begin{subequations}
\begin{align}
    \delta B^*
    &=
    -\frac{(2\pi)^2}{ik}
    \left(
        \chi_{\rho\rho}
        +
        k^2\chi_{\text{EE}}^{(0)}
        +
        k^4\chi_{\text{EE}}^{(\text{L})}(k^2)
    \right)
    \delta J_y^* ,
    \\
    \delta E_x^*
    &=
    -ik(2\pi)^2\chi_{\text{BB}}(k^2)\delta J^{t*}.
\end{align}
\end{subequations}
Varying with respect to the transformed sources gives
\begin{subequations}
\label{eq:GreensStrans}
\begin{align}
    \langle J^{y*}J^{y*}\rangle_{\mathrm R}(0,\vec k)
    &=
    - \frac{k^2}{
        (2\pi)^2
        \left(
            \chi_{\rho\rho}
            +
            k^2\chi_{\text{EE}}^{(0)}
            +
            k^4\chi_{\text{EE}}^{(\text{L})}(k^2)
        \right)
    },
    \label{eq:GyS}
    \\
    \langle J^{t*}J^{t*}\rangle_{\mathrm R}(0,\vec k)
    &=
    \frac{1}{(2\pi)^2\chi_{\text{BB}}(k^2)} .
    \label{eq:GtS}
\end{align}
\end{subequations}
This makes explicit that the $S$-transformation exchanges longitudinal and
transverse response and inverts the corresponding static susceptibilities.

The transformed microscopic theory admits an EFT of the same form,
\begin{subequations}
\label{eq:Greens_transformed}
\begin{align}
    \langle J^{t*}J^{t*}\rangle_{\mathrm R}(0,\vec k)
    &=
    -
    \left(
        \chi_{\rho\rho}^*
        +
        k^2\chi_{\text{EE}}^{(0)*}
        +
        k^4\chi_{\text{EE}}^{(\text{L})*}(k^2)
    \right),
    \label{eq:Gto}
    \\
    \langle J^{y*}J^{y*}\rangle_{\mathrm R}(0,\vec k)
    &=
    k^2\chi_{\text{BB}}^*(k^2).
    \label{eq:Gyo}
\end{align}
\end{subequations}
Matching \eqref{eq:GreensStrans} and \eqref{eq:Greens_transformed} order by
order in $k^2$ gives
\begin{align}
    \chi_{\rho\rho}^*
    &=
    -\frac{1}{(2\pi)^2\chi_{\text{BB}}(0)} ,
    \label{eq:chargesusceptibility}
    \\
    \chi_{\text{EE}}^{*(0)}
    &=
    \frac{\chi_{\text{BB}}'(0)}
    {(2\pi)^2\chi_{\text{BB}}(0)^2}.
    \label{eq:polarisationsuscS}
\end{align}
Here and below a prime denotes differentiation with respect to
\begin{equation}
    q\equiv k^2 ,
    \qquad
    \chi'(0)\equiv \left.\frac{d\chi(q)}{dq}\right|_{q=0}.
\end{equation}
With this convention the longitudinal electric susceptibility of the
transformed theory has the expansion
\begin{equation}
    \chi_{\text{EE}}^{*(\text{L})}(k^2)
    =
    \chi_{\text{EE}}^{*(\text{L})}(0)
    +
    \chi_{\text{EE}}^{*(\text{L})(1)}(0)k^2
    +
    \frac12\chi_{\text{EE}}^{*(\text{L})(2)}(0)k^4
    +
    \frac16\chi_{\text{EE}}^{*(\text{L})(3)}(0)k^6
    +
    \mathcal O(k^8),
\end{equation}
where the superscript $(i)$ denotes the $i$-th derivative with respect to
$q=k^2$. The first coefficient is
\begin{equation}
    \chi_{\text{EE}}^{*(\text{L})}(0)
    =
    \frac{
        \chi_{\text{BB}}(0)\chi_{\text{BB}}''(0)
        -
        2\chi_{\text{BB}}'(0)^2
    }
    {
        8\pi^2\chi_{\text{BB}}(0)^3
    } .
\end{equation}
Higher-order coefficients are listed in Appendix~\ref{app:mappingchargesus}.

Similarly, from \eqref{eq:GyS} and \eqref{eq:Gyo}, the transformed
magnetisation susceptibility admits the expansion
\begin{equation}
\begin{split}
    \chi_{\text{BB}}^*(k^2)
    &=
    \chi_{\text{BB}}^*(0)
    +
    \chi_{\text{BB}}^{*(1)}(0)k^2
    +
    \frac12\chi_{\text{BB}}^{*(2)}(0)k^4
    +
    \frac16\chi_{\text{BB}}^{*(3)}(0)k^6
    \\
    &\quad
    +
    \frac1{24}\chi_{\text{BB}}^{*(4)}(0)k^8
    +
    \frac1{120}\chi_{\text{BB}}^{*(5)}(0)k^{10}
    +
    \mathcal O(k^{12}) ,
\end{split}
\end{equation}
with leading coefficient
\begin{equation}
    \chi_{\text{BB}}^*(0)
    =
    - \frac{1}{4\pi^2\chi_{\rho\rho}} .
\end{equation}
The higher-order coefficients are given in Appendix~\ref{app:mappingmagsusc}.

The static susceptibilities are therefore reorganised non-trivially under the
$S$-transformation. Charge response in the transformed theory is determined
by magnetic response in the original theory, while magnetic response in the
transformed theory is determined by charge response in the original theory.

\subsubsection{Einstein relation}

Let us investigate the effect of the transformation on the Einstein relation 
which is given in the original theory by
\begin{equation}
\label{eq:Einstein}
    D=\frac{\sigma_{\mathrm{DC}}}{\chi_{\rho\rho}} .
\end{equation}
Analogously, in the transformed theory, one has
\begin{equation}
\label{eq:SEinstein}
    D^*
    =
    \frac{\sigma_{\mathrm{DC}}^*}{\chi_{\rho\rho}^*}.
\end{equation}
As discussed in Section \ref{section:S-Field}, the $S$-transformed DC conductivity is
\begin{equation}
    \sigma_{\mathrm{DC}}^*
    =
    \frac{1}{(2\pi)^2}\frac{1}{\sigma_{\mathrm{DC}}}.
\end{equation}
Using \eqref{eq:chargesusceptibility} this yields
\begin{equation}
    D^*
    =
    -\frac{\chi_{\text{BB}}(0)}{\sigma_{\mathrm{DC}}}.
\end{equation}
Thus the diffusion constant of the $S$-transformed theory is fixed by the
magnetic susceptibility and DC conductivity of the original theory. 

\subsubsection{Mapping of (quasi-)hydrodynamics}
\label{sec:quasihydro}

We now consider the finite-frequency, zero-momentum regime,
\[
    \vec k=\vec 0,
    \qquad
    \omega\neq 0 .
\]
In this limit the constitutive relation \eqref{eq:constitutive1} and the
equations of motion \eqref{eq:eom} give
\begin{equation}
\label{eq:chargecurrentk0}
    \delta J^i
    =
    \delta\bar J^i
    -
    i\omega\delta p^i ,
\end{equation}
together with
\begin{equation}
    \prod_{n=1}^{N-1}
    \left(
        -i\omega+\Gamma_{(0),n}
    \right)
    \delta\bar J^i
    =
    i\omega
    \left(
        \bar\sigma_{(0)}
        -
        i\omega r_{(0)}[-i\omega]
    \right)
    \delta a^i .
\end{equation}
Here $N-1$ is the number of retained gapped poles, while the subscript
$(0)$ denotes the $k$-independent part of the corresponding coefficient.
At $\vec k=\vec 0$, $ \omega \neq 0 $ the polarisation is
\[
    \delta p^i
    =
    i\omega\chi_{\text{EE}}^{(0)}\delta a^i ,
\]
so that
\begin{equation}
\label{eq:constitutivefinitew}
    \delta J^i
    =
    \delta\bar J^i
    -
    (i\omega)^2\chi_{\text{EE}}^{(0)}\delta a^i .
\end{equation}

For the retarded Green's function to take the desired Mittag-Leffler form, the
function $r_{(0)}(-i\omega)$ is not arbitrary. It is fixed by the AC
conductivity, the gapped pole positions, and the corresponding residues
$R_{(0),n}$:
\begin{subequations}
\begin{align}
\label{Eq:Hydrozerokconstraint}
    r_{(0)}(-i\omega)
    &=
    \frac{i}{\omega}
    \bigg[
    \left(
        \sigma_{\mathrm{AC}}(i\omega)
        +
        i\omega\chi_{\mathrm{EE}}^{(0)}
        -
        \sum_{n=0}^{\infty}\sum_{m=1}^{N-1}
        \frac{R_{(0),m}(i\omega)^n}{(\Gamma_{(0),m})^{n+1}}
    \right)
    Q(\omega)
    \nonumber
    \\
    &\qquad
    +
    \sum_{n=1}^{N-1}R_{(0),n}P_n(\omega)
    -
    \bar\sigma_{(0)}
    \bigg],
    \\
    P_n(\omega)
    &=
    \prod_{\substack{m=1 \\ m\neq n}}^{N-1}
    \left(
        -i\omega+\Gamma_{(0),m}
    \right),
    \qquad
    Q(\omega)
    =
    \prod_{n=1}^{N-1}
    \left(
        -i\omega+\Gamma_{(0),n}
    \right).
\end{align}
\end{subequations}
The apparent factor $1/\omega$ is nonsingular once the constraint is imposed, with 	
	\begin{displaymath}
		r_{(0)}(-i\omega)=O(\omega^0)
	\end{displaymath}
at small frequency. We first consider pure hydrodynamics and then include gapped poles.

\paragraph{Pure hydrodynamics.}
Consider first the purely hydrodynamic regime, in which no gapped poles
are retained in either theory. This is the case in which the duality descends to the
effective theory in the most transparent way: since no pole data have to be tracked,
the $S$-transformation maps the finite set of hydrodynamic transport coefficients of
the original theory directly onto those of the dual theory. We show below that the
map closes, i.e. every dual coefficient is determined by original-theory data alone.
The constitutive relation reduces to
\begin{equation}
    \delta J^i
    =
    i\omega
    \left(
        \bar\sigma_{(0)}
        -
        i\omega r_{(0)}[-i\omega]
        -
        i\omega\chi_{\text{EE}}^{(0)}
    \right)
    \delta a^i .
\end{equation}
Using $\delta E^i=i\omega\delta a^i$ and the pure $S$-transformation,
this can be written as
\begin{equation}
\label{eq:constitutiveoomega}
    \delta E^{*i}
    =
    (2\pi)^2
    \left(
        \bar\sigma_{(0)}
        -
        i\omega r_{(0)}[-i\omega]
        -
        i\omega\chi_{\text{EE}}^{(0)}
    \right)
    J^{*i}.
\end{equation}
The transformed EFT without gapped poles has the form
\begin{equation}
\label{eq:constitutivetomega}
    \delta J^{i*}
    =
    i\omega
    \left(
        \bar\sigma_{(0)}^*
        -
        i\omega r_{(0)}^*[-i\omega]
        -
        i\omega\chi_{\text{EE}}^{(0)*}
    \right)
    \delta a^{i*}.
\end{equation}
Expanding the corresponding Green's functions at small $\omega$ and matching
order by order gives
\begin{equation}
    \bar\sigma_{(0)}^*
    =
    \frac{1}{(2\pi)^2\bar\sigma_{(0)}} .
\end{equation}
The function $r_{(0)}^*(-i\omega)$ has the expansion
\begin{equation}
    r_{(0)}^*(-i\omega)
    =
    r_{(0)}^*(0)
    +
    r_{(0)}^{*(1)}(0)\omega
    +
    \frac12 r_{(0)}^{*(2)}(0)\omega^2
    +
    \mathcal O(\omega^3),
\end{equation}
with leading coefficient
\begin{equation}
    r_{(0)}^*(0)
    =
    -
    \frac{
        r_{(0)}(0)+\chi_{\text{EE}}^{(0)}
    }
    {
        (2\pi)^2\bar\sigma_{(0)}^2
    }
    -
    \chi_{\text{EE}}^{*(0)} .
\end{equation}
Using \eqref{eq:polarisationsuscS}, this becomes
\begin{equation}
    r_{(0)}^*(0)
    =
    -
    \frac{1}{(2\pi)^2}
    \left[
        \frac{r_{(0)}(0)+\chi_{\text{EE}}^{(0)}}
        {\bar\sigma_{(0)}^2}
        +
        \frac{\chi_{\text{BB}}'(0)}
        {\chi_{\text{BB}}(0)^2}
    \right].
\end{equation}
In the purely hydrodynamic regime, therefore, the transformed transport
coefficients are determined entirely by those of the original theory.
This is the sharpest positive statement of the paper: through the pure
hydrodynamic sector the duality acts as a genuine transformation of the effective
theory, mapping a finite list of original transport coefficients onto the dual list
without any auxiliary pole data. The complication analysed below arises only when
long-lived non-hydrodynamic modes are present, and even then it is resolved by
supplementing this map with the transformed pole data.

\paragraph{Including gapped poles.}
The situation changes once gapped poles are included. The number of gapped
poles that must be retained in the transformed EFT is determined by the pole
structure of the transformed microscopic correlator inside the chosen domain.
In the D3/D5 application discussed in Section~\ref{sec:D3D5}, the relevant
domain contains three different regimes: one pole in the original theory may
map to zero, one, or two gapped poles in the transformed theory.
In each of these regimes the dictionary still closes: the dual
coefficients are given explicitly in terms of the original transport data together
with the positions and residues of the transformed poles. What changes across
regimes is only \emph{which} pole data must be supplied, not whether an explicit map
exists.

For $N-1$ gapped poles, the zero-momentum constitutive relation becomes
\begin{equation}
    \delta J^i
    =
    i\omega
    \left(
        \frac{
            \bar\sigma_{(0)}
            -
            i\omega r_{(0)}[-i\omega]
            -
            i\omega
            \prod_{n=1}^{N-1}
            \left(
                -i\omega+\Gamma_{(0),n}
            \right)
            \chi_{\text{EE}}^{(0)}
        }
        {
            \prod_{n=1}^{N-1}
            \left(
                -i\omega+\Gamma_{(0),n}
            \right)
        }
    \right)
    \delta a^i .
\end{equation}

As the simplest non-trivial example, consider the region where the original EFT
contains one gapped pole, while the transformed EFT contains two. The original
constitutive relation is
\begin{equation}
\label{eq:const1pole}
    \delta J^i
    =
    i\omega
    \left(
        \frac{
            \bar\sigma_{(0)}
            -
            i\omega r_{(0)}[-i\omega]
            -
            i\omega
            \left(
                -i\omega+\Gamma_{(0),1}
            \right)
            \chi_{\text{EE}}^{(0)}
        }
        {
            -i\omega+\Gamma_{(0),1}
        }
    \right)
    \delta a^i .
\end{equation}
The transformed EFT with two gapped poles is
\begin{equation}
\label{eq:constitutive_finitew}
    \delta J^{i*}
    =
    i\omega
    \frac{
        \bar\sigma_{(0)}^*
        -
        i\omega r_{(0)}^*[-i\omega]
        -
        i\omega
        \left(
            -i\omega+\Gamma_{(0),1}^*
        \right)
        \left(
            -i\omega+\Gamma_{(0),2}^*
        \right)
        \chi_{\text{EE}}^{(0)*}
    }
    {
        \left(
            -i\omega+\Gamma_{(0),1}^*
        \right)
        \left(
            -i\omega+\Gamma_{(0),2}^*
        \right)
    }
    \delta a^{i*}.
\end{equation}
Matching the leading term gives
\begin{equation}
    \bar\sigma_{(0)}^*
    =
    \frac{
        \Gamma_{(0),1}
        \Gamma_{(0),1}^*
        \Gamma_{(0),2}^*
    }
    {
        (2\pi)^2\bar\sigma_{(0)}
    } .
\end{equation}
This expression already shows the key point: in this regime
$\bar\sigma_{(0)}^*$ cannot be fixed from the original EFT coefficients
alone. The pole positions of the transformed correlator must be supplied as
additional EFT data. Equivalently, the transformed pole positions are
non-perturbative data from the point of view of a finite low-frequency
expansion of the original correlator.

For the leading coefficient of $r_{(0)}^*(-i\omega)$ one finds
\begin{equation}
\begin{split}
    r_{(0)}^*(0)
    =
    &\biggl[
    \Gamma_{(0),1}^*
    \biggl(
    -
    \Gamma_{(0),1}\Gamma_{(0),2}^*
    \left(
        \Gamma_{(0),1}\chi_{\text{EE}}^{(0)}
        +
        r_{(0)}(0)
    \right)
    -
    \bar\sigma_{(0)}^2
    \frac{\chi_{\text{BB}}'(0)}
    {\chi_{\text{BB}}(0)^2}
    \Gamma_{(0),2}^*
    \\
    &\qquad
    +
    \bar\sigma_{(0)}
    \left(
        \Gamma_{(0),1}
        +
        \Gamma_{(0),2}^*
    \right)
    \biggr)
    +
    \Gamma_{(0),1}\Gamma_{(0),2}^*
    \bar\sigma_{(0)}
    \biggr]
    \frac{1}
    {
        (2\pi)^2\bar\sigma_{(0)}^2
    } .
\end{split}
\end{equation}

The one-to-one and one-to-zero maps are given in
Appendix~\ref{app:coefffinite}. When the transformed EFT retains one or more
gapped poles, their positions, and in general also their residues, must be supplied
as transformed pole data. The one-to-zero regime is the exception: since no gapped
pole is retained on the transformed side, no transformed pole data are needed.

For numerical purposes it is often more convenient to use the
Mittag-Leffler form of the zero-momentum current-current correlator,
\begin{subequations}
\label{eq:currentCorrelatorZerok2}
\begin{align}
    \langle J^iJ^j\rangle_{\mathrm R}(\omega,\vec 0)
    &=
    -i\omega
    \left[
        \sum_{n=0}^{N_D}c_n(i\omega)^n
        +
        \sum_{n=1}^{N-1}
        \frac{iR_{(0),n}}{\omega+i\Gamma_{(0),n}}
    \right]\delta^{ij}
    +
    \mathcal O(\omega^{N_D+1}),
    \\
    c_n
    &=
    \frac{1}{n!}\sigma_{\mathrm{AC}}^{(n)}(0)
    -
    \sum_{m=1}^{N-1}
    \frac{R_{(0),m}}{(\Gamma_{(0),m})^{n+1}} . \label{eq:coefficients}
\end{align}
\end{subequations}
Repeating the matching directly in this representation gives the leading
coefficient $c_0^*$. For a one-to-one map,
\begin{equation}
    c_0^*
    =
    \frac{\Gamma_{(0),1}}
    {
        (2\pi)^2
        \left(
            c_0\Gamma_{(0),1}
            +
            R_{(0),1}
        \right)
    }
    -
    \frac{R_{(0),1}^*}{\Gamma_{(0),1}^*}.
\end{equation}
For a one-to-two map,
\begin{equation}
    c_0^*
    =
    \frac{\Gamma_{(0),1}}
    {
        (2\pi)^2
        \left(
            c_0\Gamma_{(0),1}
            +
            R_{(0),1}
        \right)
    }
    -
    \frac{R_{(0),1}^*}{\Gamma_{(0),1}^*}
    -
    \frac{R_{(0),2}^*}{\Gamma_{(0),2}^*}.
\end{equation}
For a one-to-zero map,
\begin{equation}
    c_0^*
    =
    \frac{\Gamma_{(0),1}}
    {
        (2\pi)^2
        \left(
            c_0\Gamma_{(0),1}
            +
            R_{(0),1}
        \right)
    } .
\end{equation}
Thus the transformed holomorphic coefficient depends on the pole data retained
in the transformed EFT. The higher-order coefficients are listed in
Appendix~\ref{app:coefffinite}. The numerical extraction of these coefficients
for the D3/D5 system is discussed in Section~\ref{sec:D3D5}.

\subsection{\texorpdfstring{$SL(2,\mathbb Z)$-transformed theory}{SL(2,Z)-transformed theory}}\label{Sec:SL2Zgeneral}

The previous subsections focused on the pure $S$-transformation. A general
$SL(2,\mathbb Z)$ transformation can be treated similarly, using the field
and current transformations in \eqref{eq:SL-Field-Transformations} and
\eqref{eq:Field-SL-Transformations}. Since the holographic application studied
below only requires the pure $S$-transformation, we restrict the discussion
here to the static susceptibility sector.

At $\omega=0$, the original constitutive relations become, after using the
inverse $SL(2,\mathbb Z)$ transformation,
\begin{subequations}
\begin{align}
    d\,\delta J^{t*}
    +
    \frac{b}{2\pi}\delta B^*
    &=
    \left(
        \chi_{\rho\rho}
        +
        k^2\chi_{\text{EE}}^{(0)}
        +
        k^4\chi_{\text{EE}}^{(\text{L})}(k^2)
    \right)
    \frac{1}{ik}
    \left(
        a\,\delta E_x^*
        +
        2\pi c\,\delta J_y^*
    \right),
    \\
    d\,\delta J^{y*}
    +
    \frac{b}{2\pi}\delta E_x^*
    &=
    -
    k^2\chi_{\text{BB}}(k^2)
    \frac{
        a\,\delta B^*
        +
        2\pi c\,\delta J^{t*}
    }
    {ik}.
\end{align}
\end{subequations}
Introducing
\[
    \tilde\chi_{\text{EE}}(k^2)
    =
    \chi_{\rho\rho}
    +
    k^2\chi_{\text{EE}}^{(0)}
    +
    k^4\chi_{\text{EE}}^{(\text{L})}(k^2),
    \qquad
    \tilde\chi_{\text{BB}}(k^2)
    =
    \chi_{\text{BB}}(k^2),
\]
one obtains
\begin{subequations}
\begin{align}
    \delta J^{t*}
    &=
    -
    \frac{1}{ik}
    \frac{
        \tilde\chi_{\text{EE}}
    }
    {
        c^2(2\pi)^2
        \tilde\chi_{\text{BB}}\tilde\chi_{\text{EE}}
        -
        d^2
    }
    \delta E_x^*
    \nonumber
    \\
    &\quad
    +
    \frac{1}{2\pi}
    \frac{
        bd
        -
        ac(2\pi)^2
        \tilde\chi_{\text{BB}}\tilde\chi_{\text{EE}}
    }
    {
        c^2(2\pi)^2
        \tilde\chi_{\text{BB}}\tilde\chi_{\text{EE}}
        -
        d^2
    }
    \delta B^* ,
    \\
    \delta J_y^*
    &=
    -
    ik
    \frac{
        \tilde\chi_{\text{BB}}
    }
    {
        c^2(2\pi)^2
        \tilde\chi_{\text{BB}}\tilde\chi_{\text{EE}}
        -
        d^2
    }
    \delta B^*
    \nonumber
    \\
    &\quad
    +
    \frac{1}{(2\pi)^2}
    \frac{
        bd
        -
        ac(2\pi)^2
        \tilde\chi_{\text{BB}}\tilde\chi_{\text{EE}}
    }
    {
        c^2(2\pi)^2
        \tilde\chi_{\text{BB}}\tilde\chi_{\text{EE}}
        -
        d^2
    }
    \delta E_x^* .
\end{align}
\end{subequations}
Varying with respect to the transformed sources gives
\begin{subequations}
\begin{align}
    \langle J^{y*}J^{y*} \rangle_{\text{R}}(0,\vec{k})
    &=
    -
    k^2
    \frac{
        \tilde\chi_{\text{BB}}
    }
    {
        c^2(2\pi)^2
        \tilde\chi_{\text{BB}}\tilde\chi_{\text{EE}}
        -
        d^2
    } , \\ 
    \langle J^{t*}J^{t*}\rangle_{\text{R}}(0,\vec{k})
    &=
    \frac{
        \tilde\chi_{\text{EE}}
    }
    {
        c^2(2\pi)^2
        \tilde\chi_{\text{BB}}\tilde\chi_{\text{EE}}
        -
        d^2
    } .
\end{align}
\end{subequations}

\noindent Matching to the transformed EFT gives, at leading orders,
\begin{subequations}
\begin{align}
    \chi_{\rho\rho}^*
    &=
    -
    \frac{\chi_{\rho\rho}}
    {
        (2\pi)^2c^2\chi_{\text{BB}}(0)\chi_{\rho\rho}
        -
        d^2
    },
    \\
    \chi_{\text{EE}}^{*(0)}
    &=
    \frac{
        (2\pi)^2c^2\chi_{\rho\rho}^2\chi_{\text{BB}}'(0)
        +
        d^2\chi_{\text{EE}}^{(0)}
    }
    {
        \left(
            d^2
            -
            (2\pi)^2c^2\chi_{\text{BB}}(0)\chi_{\rho\rho}
        \right)^2
    } .
\end{align}
\end{subequations}
For other regimes analogous relations can be derived. However, since they are not relevant to us, we do not explicitly state them here. Setting $a=d=0 $ and $b=-c=1$, the mappings reduce to those of the $S$-transformation computed in Section \ref{sec:suscmapping}, as expected.

\section{Alternative quantised D3/D5 brane system}
\label{sec:D3D5}

In this section we illustrate the mechanism developed above in the holographic D3/D5 probe brane system at finite temperature, parametrised by the standard-quantisation charge density $\rho$~\cite{Brattan:2013wya,Brattan:2014moa,Jokela:2014wsa,Itsios:2016ffv,Jokela:2017fwa}. The model provides a controlled laboratory in which the analytic mappings derived in Section~\ref{sec:Stransformation} can be tested explicitly. The field-theory $S$-operation is implemented holographically by changing the boundary condition of the worldvolume gauge field, or equivalently by passing from standard to alternative quantisation, which in this probe-brane setting realises a holographic version of the anyonic $S$-duality~\cite{Brattan:2013wya,Brattan:2014moa}.

It is important to emphasise that the D3/D5 probe brane system considered here is not, by itself, $S$-invariant. The $S$-transformation is therefore not a symmetry of the probe-brane action, and standard and alternative quantisation should not be regarded as two descriptions of the same boundary theory. Rather, they define two distinct boundary theories whose current correlation functions are related by the field-theory $S$-transformation reviewed in Section~\ref{sec:Stransformation}.

We use the D3/D5 system not as an example of an $S$-self-dual theory, but as a holographic setup in which both sides of the transformation can be computed explicitly. The question is whether the low-frequency effective theory reconstructed from the transformed correlator agrees with the result of naively $S$-transforming the low-frequency effective theory of the original system. As discussed above, these operations do not commute in general: the $S$-transformation acts on the full meromorphic response function and can therefore change the set of poles lying inside the effective domain.

With this said, throughout this section, $\rho$ denotes the electric-flux parameter of the bulk solution, which in standard quantisation is the charge density. The standard-quantisation background considered below has $B=0$, and therefore the pure $S$-transformation, ~\eqref{eq:SL-Field-Transformations} with $a = d = 0$ and $b = - c = 1$, gives
\begin{equation}
    \rho^*=0,
    \qquad
    B^*=2\pi\rho .
\end{equation}
Thus, in the alternatively quantised theory, varying $\rho$ should equivalently be understood as varying the transformed background magnetic field $B^*$ rather than the transformed charge density. In terms of the dimensionless variables introduced below, $\widetilde B^*=2\pi\widetilde\rho$. Although the transformed background has $B^*\neq0$, the original zero-field response has no Hall component; consequently, the pure $S$-transformation preserves the diagonal form of the zero-momentum conductivity in the state considered here, and the scalar inversion relation~\eqref{eq:S_conductivity} remains applicable.

We numerically compute the static susceptibilities, diffusion constant, and low-frequency AC conductivity of the alternatively quantised D3/D5 system and compare them with the predictions of the analytic mappings. The agreement provides a non-trivial check of the formalism and makes explicit how quasihydrodynamic transport data are reorganised under the $S$-transformation. In the large-density regime, the transformed conductivity reduces to the inverse Drude form implied by the exact transformation of the correlator. Capturing corrections to this behaviour requires enlarging the effective domain and retaining the additional pole data that enter it.

\subsection{Alternative quantisation in holography}
\label{sec:holographicSL2Z}

For a bulk gauge field in $AdS_4$, the field-theory $SL(2,\mathbb{Z})$ action on the dual current correlators can be implemented holographically through mixed boundary conditions \cite{Jokela:2013hta}. In particular, the $S$-transformation relevant for this paper is realised holographically by changing the boundary condition of the bulk gauge field, i.e. by passing from standard to alternative quantisation. In standard quantisation one imposes Dirichlet boundary conditions on the gauge field. The variation of the renormalised bulk action then contains the boundary term
\begin{equation}
    \delta S_{\mathrm D}
    =
    \int_{z=0} J^\mu \delta A_\mu ,
\end{equation}
where the conformal boundary is at $z=0$ in Eddington-Finkelstein coordinates and
\begin{equation}
    J^\mu = \frac{\delta S_{\mathrm D}}{\delta A_\mu(z=0)}
\end{equation}
is interpreted as the conserved current of the boundary theory. To introduce mixed boundary conditions one may add the boundary term
\begin{equation}
    \label{eq:Sgeneric}
    S_{\mathrm{generic}}
    =
    S_{\mathrm D}
    +
    \frac{1}{2\pi}
    \int_{\mathrm{boundary}}
    \left[
        a_1 \epsilon_{\mu\rho\nu} A^\mu \partial^\rho v^\nu
        +a_2 \epsilon_{\mu\rho\nu} A^\mu \partial^\rho A^\nu
        +a_3 \epsilon_{\mu\rho\nu} v^\mu \partial^\rho v^\nu
    \right] .
\end{equation}
Here we have used current conservation to write
\begin{equation}
    J^\mu = \frac{1}{2\pi}\epsilon^{\mu\nu\rho}\partial_\rho v_\nu ,
\end{equation}
with $v_\nu$ defined up to gauge transformations. Defining also
\begin{equation}
    \mathcal B_\mu
    \equiv
    \frac{1}{2\pi}\epsilon_{\mu\nu\rho}\partial^\nu A^\rho,
\end{equation}
the variation can be written as
\begin{equation}
    \delta S_{\mathrm{generic}}
    =
    \int_{\mathrm{boundary}}
    \left(aJ_\mu+b\mathcal B_\mu\right)
    \left(c\delta v^\mu+d\delta A^\mu\right) ,
\end{equation}
where
\begin{equation}
    ad=1+a_1,
    \qquad
    bc=a_1,
    \qquad
    bd=2a_2,
    \qquad
    ac=2a_3 .
\end{equation}
Since $ad-bc=1$, an arbitrary change of boundary conditions corresponds to an $SL(2,\mathbb{R})$ transformation. At the level of correlation functions this continuous group acts on the bulk gauge field, and the alternative quantisation used below realises the specific $S$-element with $a=d=0$, $b=-c=1$. The restriction to the arithmetic subgroup $SL(2,\mathbb{Z})$ is a statement about the boundary theory rather than about the bulk: it arises once one demands that the added boundary terms in~\eqref{eq:Sgeneric} define a consistent, gauge-invariant partition function on a general closed three-manifold. The Chern-Simons contact term $a_2\,\epsilon_{\mu\rho\nu}A^\mu\partial^\rho A^\nu$ (equivalently the $T$-transformation) is well defined only for integer level, and combined with the gauging operation $S$ this quantises $a,b,c,d\in\mathbb{Z}$~\cite{Witten:2003ya,Leigh:2003ez}. In the normalisation adopted here, with the magnetic current defined through $\tfrac{1}{2\pi}\epsilon^{\mu\nu\rho}\partial_\nu A_\rho$, the level appearing in~\eqref{eq:Sgeneric} is $a_2\in\mathbb{Z}$, so that the physically distinct boundary conditions are labelled by $SL(2,\mathbb{Z})$. For the linear-response quantities computed in this paper, which only involve the pure $S$-operation on two-point functions, this arithmetic restriction plays no role, and the continuous relation~\eqref{eq:DualConductivity} suffices.

The new boundary condition fixes the combination $cv^\mu+dA^\mu$, or equivalently the gauge-invariant quantity
\begin{equation}
    \delta \mathcal B^*_\mu
    =
    c\delta J_\mu+d\delta\mathcal B_\mu
    =0 .
\end{equation}
The corresponding current of the transformed boundary theory is
\begin{equation}
    J^*_\mu = aJ_\mu+b\mathcal B_\mu .
\end{equation}
Thus the transformed and original variables are related by
\begin{equation}
    \begin{pmatrix}
        J^*_\mu \\
        \mathcal B^*_\mu
    \end{pmatrix}
    =
    \begin{pmatrix}
        a & b \\
        c & d
    \end{pmatrix}
    \begin{pmatrix}
        J_\mu \\
        \mathcal B_\mu
    \end{pmatrix} .
\end{equation}
The pure $S$-transformation corresponds to $a=d=0$ and $b=-c=1$.

\subsection{Fluctuations of the D3/D5 brane model}

We now apply the mappings to the alternatively quantised D3/D5 brane system at non-zero temperature on the background parameterised by $\rho$. Let us first briefly review the model and the equations of motion for linearised charge fluctuations. A more detailed review can be found in Ref.~\cite{Amoretti:2025kem}. Working in ingoing Eddington-Finkelstein coordinates, the geometry generated by the backreacted D3-branes is
\begin{subequations}
\label{Eq:spacetime}
\begin{align}
    ds^2
    &=
    \frac{L^2}{r^2}
    \left[-f(r)dv^2-2dvdr+dx^2+dy^2+dz^2\right]
    +L^2 ds^2_{S^5},
    \\
    f(r)
    &=
    1-\frac{r^4}{r_H^4} .
\end{align}
\end{subequations}
Here $r_H=(\pi T)^{-1}$, with $T$ the Hawking temperature, and $L$ is the AdS radius, which we set to $L=1$. In the equations below we measure the radial coordinate in units of $r_H$, so that the conformal boundary is at $r=0$ and the black-hole horizon is at $r_H=1$. In the probe limit the $N_f$ D5-branes do not backreact on the black-hole geometry, and their embedding is described by the Dirac-Born-Infeld action
\begin{equation}
\label{Eq:D3D5probe}
    S_{\mathrm{D5}}
    =
    -N_f T_{\mathrm{D5}}
    \int d^6\xi\,
    \sqrt{-\det\left(g_{ab}+F_{ab}\right)} .
\end{equation}
Here $\xi$ are the embedding coordinates of the D5-brane in the ten-dimensional spacetime~\eqref{Eq:spacetime}, $T_{\mathrm{D5}}$ is the D5-brane tension, $g_{ab}$ is the induced worldvolume metric, and $F_{ab}$ is the $U(1)$ worldvolume field strength. We have absorbed a factor of $2\pi\alpha'$, with $\alpha'$ the string tension, into the field strength, so that $F_{ab}$ is dimensionless.

In standard quantisation, we consider a non-zero charge density generated by a non-trivial profile $A_v(r)$, with worldvolume electric field $F_{rv}(r)=A'_v(r)$. The D5-brane action then becomes
\begin{equation}
\label{eq:action}
    S_{\mathrm{D5}}^{(0)}
    =
    -\mathcal N_5 V_{\mathbb R^{(2,1)}}
    \int_{0}^{1}dr\,
    \frac{\sqrt{1-r^4 A_v^{\prime 2}}}{r^4} ,
\end{equation}
where $\mathcal N_5\equiv N_fT_{\mathrm{D5}}V_{S^2}$, $V_{S^2}$ is the volume of the unit two-sphere, and $V_{\mathbb R^{(2,1)}}$ is the volume of the boundary spacetime.

To compute the two-point functions of charge currents we solve the linearised equations of motion for the fluctuations of the probe-brane gauge field. As in the effective theory, we align the spatial momentum along the $x$-direction and allow the fluctuations to depend only on $r$, $t$, and $x$. We use the Fourier convention
\begin{equation}
    \delta a_\mu(r,x^\mu)
    =
    \int \frac{d\omega\,dk}{(2\pi)^2}
    \delta a_\mu(r,\omega,k)
    \exp(-i\omega t+ikx) .
\end{equation}
By spatial rotation invariance, the two gauge-invariant combinations are
\begin{equation}
    \delta a_y(r,\omega,k),
    \qquad
    E_x(r,\omega,k)
    \equiv
    k\,\delta a_v(r,\omega,k)
    +
    \omega\,\delta a_x(r,\omega,k) .
\end{equation}
The corresponding equations of motion are
\begin{subequations}
\label{Eq:FluctuationEquations}
\begin{align}
\label{eq:eomEx}
E_x''
&+
\frac{
    f(r)
    \left[
        \omega^2\left(f'(r)+2i\omega\right)
        +2\omega u(r)^2\left(\rho^2r^3\omega-ik^2\right)
        -6k^2\rho^2r^3u(r)^4
    \right]
}{
    i\omega u(r)^2\left(k^2f'(r)+\omega\left(2\rho^2r^3\omega+ik^2\right)\right)
    +u(r)^4\left(k^4-6ik^2\rho^2r^3\omega\right)
}
E_x'
\nonumber \\[4pt]
&+
\frac{
    if(r)^2\left(\omega^2-k^2u(r)^2\right)
}{
    u(r)^4\left(6k^2\rho^2r^3\omega+ik^4\right)
    -\omega u(r)^2\left(k^2f'(r)+\omega\left(2\rho^2r^3\omega+ik^2\right)\right)
}
E_x=0, \\[8pt]
\label{eq:eomAy}
\delta a_y''
&+
\frac{
    u(r)^2
    \left[
        2r^3\left(-2+\rho^2\left(-3r^4+ir\omega+1\right)\right)+2i\omega
    \right]
}{f(r)^2}
\delta a_y' \nonumber \\ &
-
\frac{u(r)^2\left(k^2-2i\rho^2r^3\omega\right)}{f(r)^2}
\delta a_y=0,
\end{align}
\end{subequations}
where
\begin{equation}
    u(r)^2
    \equiv
    \frac{1-r^4}{1+r^4\rho^2},
\end{equation}
and the background profile obeys
\begin{equation}
    A_v'(r)
    =
    \frac{\rho}{\sqrt{1+r^4\rho^2}} .
\end{equation}

At $k=0$, the bulk equations for $E_x$ and $a_y$ decouple and reduce to the same equation. In what follows we use temperature-normalised variables,
\begin{equation}
    \tilde k=\frac{k}{\pi T},
    \qquad
    \tilde\omega=\frac{\omega}{\pi T},
    \qquad
    \tilde\rho=\frac{\rho}{(\pi T)^2} .
\end{equation}
The fluctuation equations can be solved numerically either by a shooting method or by the holographic-approximant method of Ref.~\cite{Amoretti:2025kem}, imposing analyticity at the horizon $r=r_H\equiv1$.

For the $S$-transformed theory, analytic results for the longitudinal DC conductivity and charge susceptibility are known~\cite{Brattan:2013wya,Brattan:2014moa}. In the current normalisation used throughout this paper they are
\begin{subequations}
\begin{align}
    \chi^*_{\tilde\rho\tilde\rho}
    &=
    \frac{1}{(2\pi)^2\,{}_2F_1\left(\frac14,\frac12,\frac54;-\tilde\rho^2\right)}
    \label{eq:analyticchargesus}
    \\
    &\overset{\tilde\rho\gg1}{=}
    \frac{\sqrt{\tilde\rho}}{(2\pi)^2}
    \left[
        \frac{\sqrt{16\pi}}{\Gamma\left(\frac14\right)^2}
        +
        \frac{16\pi}{\Gamma\left(\frac14\right)^4\sqrt{\tilde\rho}}
        +
        \mathcal O\left(\tilde\rho^{-3/2}\right)
    \right],
    \\
    \sigma^*_{(L)}
    &=
    \frac{1}{(2\pi)^2\sqrt{1+\tilde\rho^2}}
    \\
    &\overset{\tilde\rho\gg1}{=}
    \frac{1}{\tilde\rho}
    \left[
        \frac{1}{(2\pi)^2}
        -
        \frac{1}{2(2\pi)^2\tilde\rho^2}
        +
        \mathcal O\left(\tilde\rho^{-3}\right)
    \right] .
\end{align}
\end{subequations}
The Einstein relation then gives
\begin{equation}
\label{eq:Sdiffusion}
    \tilde D^*
    =
    \frac{1}{\sqrt{\tilde\rho^(2+1)}}
    {}_2F_1\left(\frac14,\frac12,\frac54;-\tilde\rho^2\right),
\end{equation}
in units where $\pi T=1$ and $L=1$.

\subsection{Susceptibilities at non-zero wavevector}
\label{sec:D3D5susc}

\begin{figure}[t!]
  \centering
  \begin{subfigure}{0.45\textwidth}
    \centering
    \includegraphics[width=\linewidth]{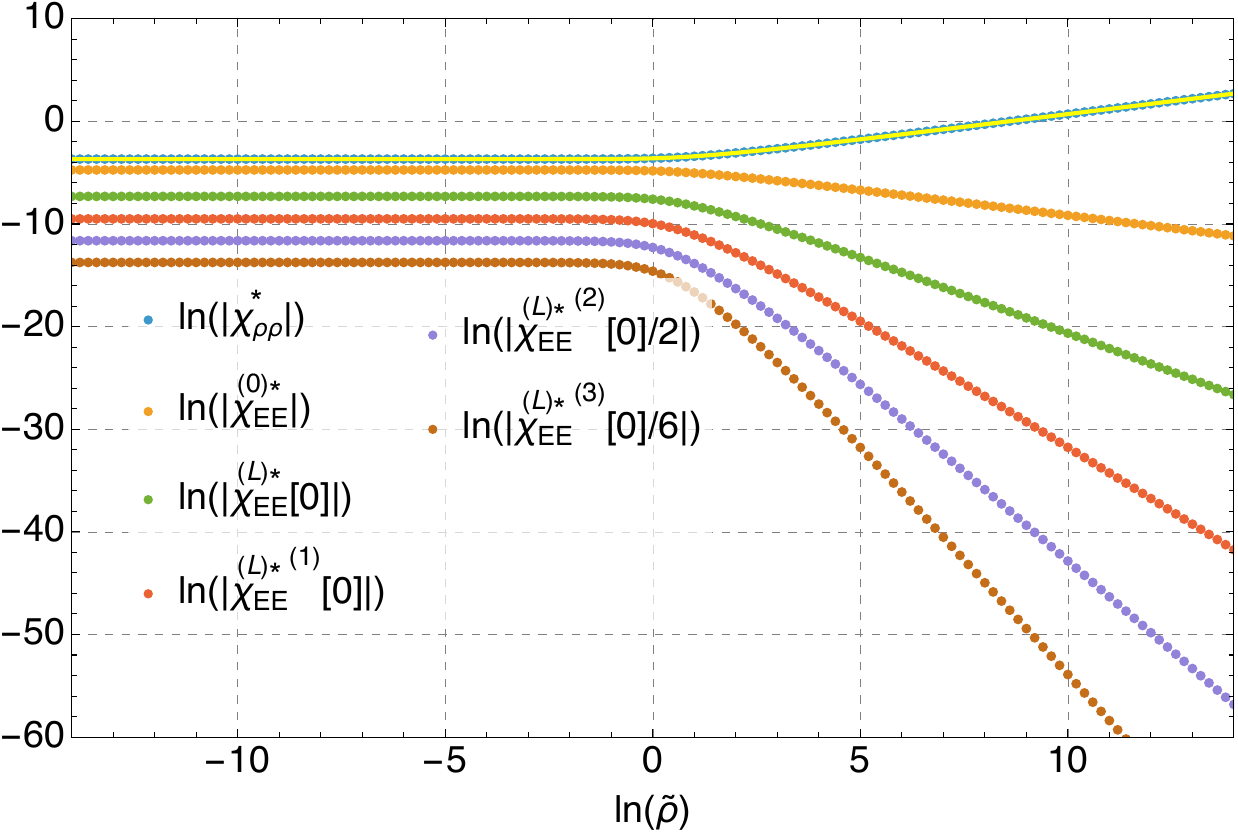}
    \caption{}
    \label{fig:dual_charge_susceptibility}
  \end{subfigure}
  \hfill
  \begin{subfigure}{0.45\textwidth}
    \centering
    \includegraphics[width=\linewidth]{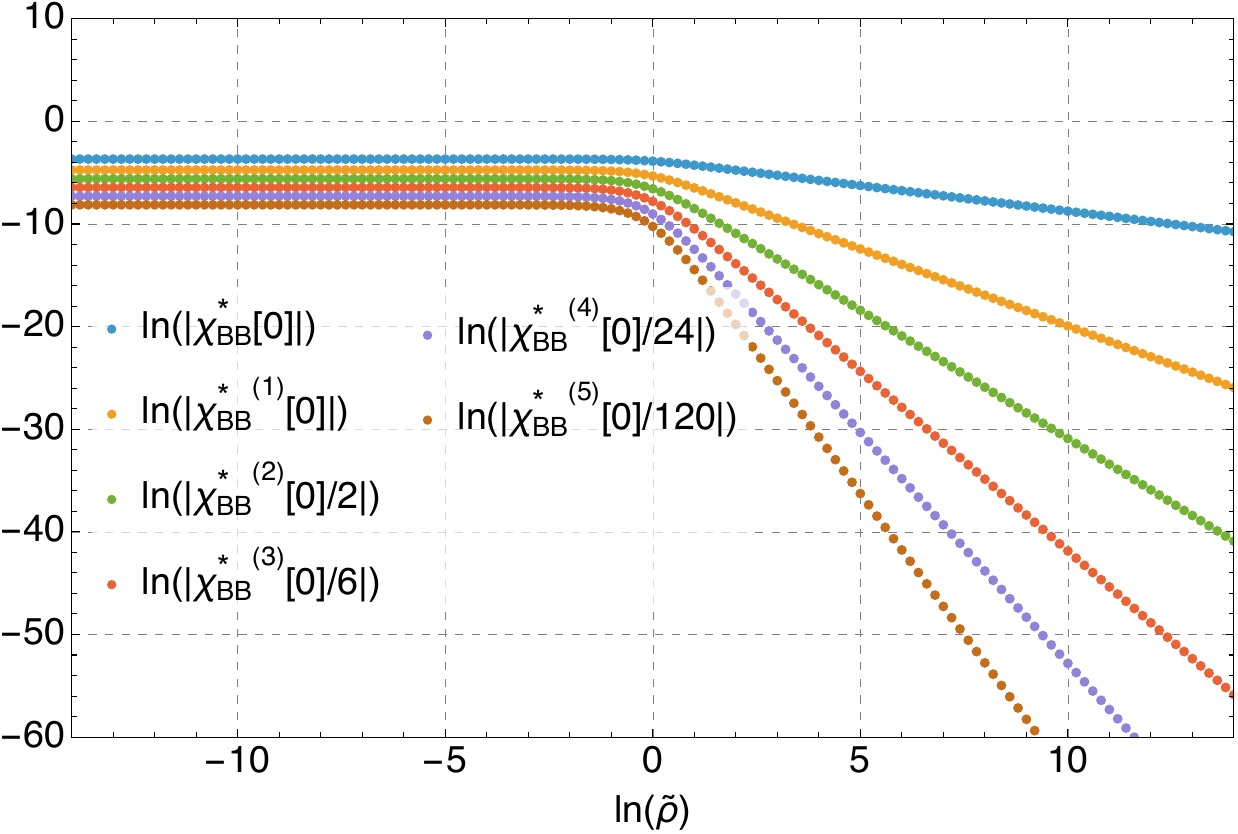}
    \caption{}
    \label{fig:dual_magnetic_susceptibility}
  \end{subfigure}
  \caption{First coefficients of the small-$\tilde{k}^2$ expansion of (a) the charge susceptibility and (b) the magnetic susceptibility as functions of $\tilde{\rho}$. At large $\tilde{\rho}$, the coefficients of the charge susceptibility scale as $\tilde\rho^{1/2-n}$, as in the ordinary quantisation but with a different proportionality constant. The analytic expression~\eqref{eq:analyticchargesus} agrees with the leading coefficient of the small-$\tilde{k}^2$ expansion. The coefficients of the magnetisation susceptibility scale as $\tilde\rho^{-1/2-n}$.}
  \label{fig:dual_susceptibilities}
\end{figure}

We now present the numerically obtained susceptibilities $\chi^*_{\rho\rho}$, $\chi^*_{\mathrm{EE}}(\vec k)$, and $\chi^*_{\mathrm{BB}}(\vec k)$ of the alternatively quantised system, i.e. of the theory related to the original one by the $S$-transformation. Figure~\ref{fig:dual_susceptibilities} shows the first coefficients of the small-$\tilde k^2$ expansion of the charge and magnetic susceptibilities as functions of charge density.

As in the ordinary quantised system, higher-order terms in the momentum expansion of the charge susceptibility are suppressed relative to the leading coefficient at large density. The leading analytic result~\eqref{eq:analyticchargesus} grows as $\sqrt{\tilde\rho}$ and agrees with the numerical extraction shown in Figure~\ref{fig:dual_charge_susceptibility}. Figure~\ref{fig:dual_magnetic_susceptibility} shows that the coefficients of the magnetic susceptibility instead scale as $\tilde\rho^{-1/2-n}$. We have also checked that the analytic susceptibility maps derived in Section~\ref{sec:suscmapping} reproduce the numerical results of the alternatively quantised system.

\begin{figure}
    \centering
    \includegraphics[width=0.5\linewidth]{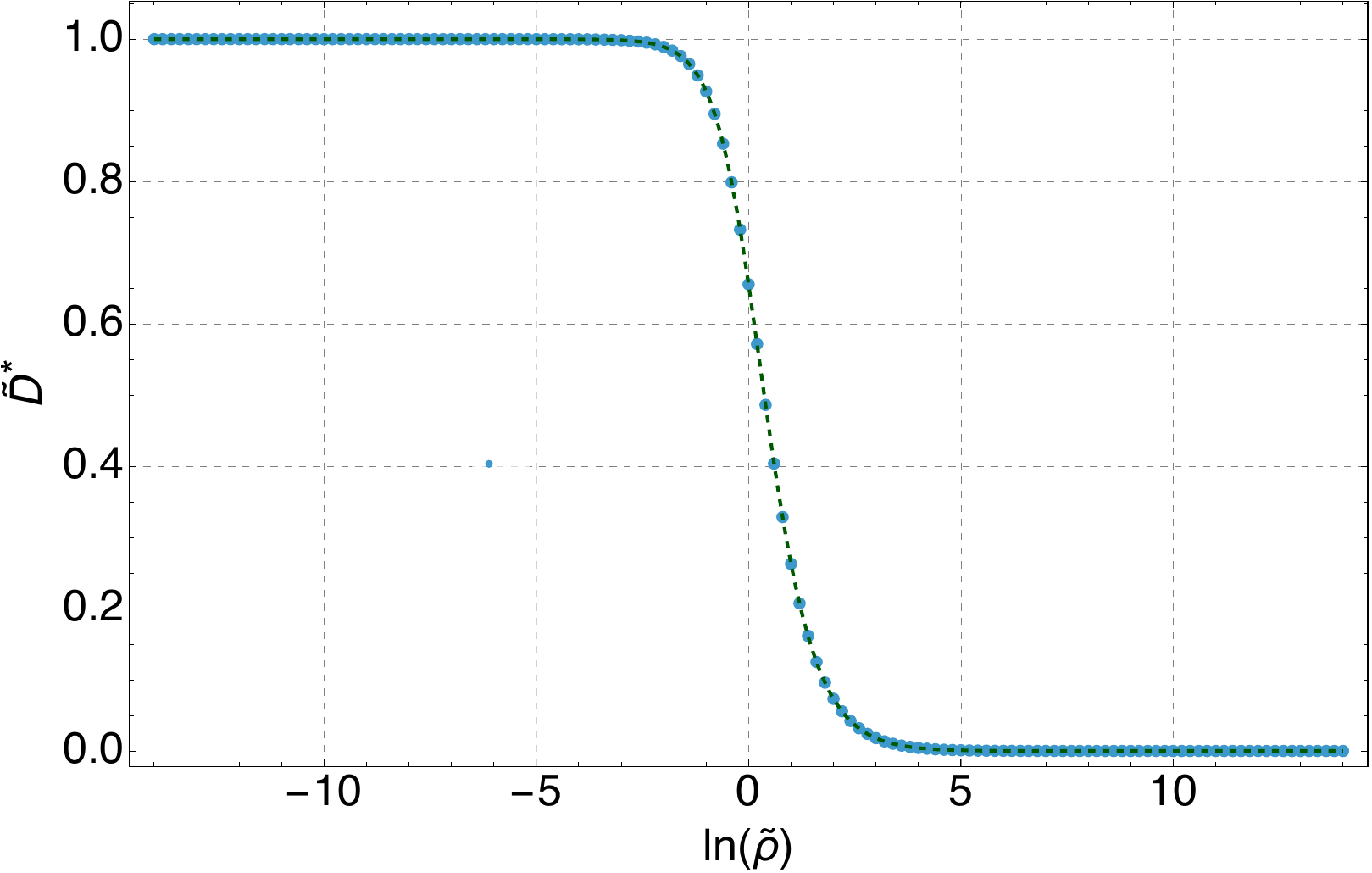}
    \caption{The dashed green curve shows the analytic expression for the $S$-transformed diffusion constant~\eqref{eq:Sdiffusion}, while the blue points show the numerical result.}
    \label{fig:diffusion}
\end{figure}

Figure~\ref{fig:diffusion} shows the diffusion constant. The green dashed curve is the analytic result~\eqref{eq:Sdiffusion}, while the blue points are obtained numerically using the magnetisation susceptibility, the Einstein relation, and the DC conductivity.

\subsection{Conductivities}

We finally consider the zero-momentum limit, $\vec k\to\vec0$. In this regime the AC conductivity of the transformed theory is written in Mittag-Leffler form as
\begin{align}
\label{eq:ACeffective}
    \sigma^*_{\mathrm{AC}}(i\omega)
    &=
    \sum_{p=1}^{N^*_{\mathrm{gap}}}
    \frac{iR^*_{(0),p}}{\omega+i\Gamma^*_{(0),p}}
    +
    \sum_{n=0}^{N_D} c^*_n(i\omega)^n
    +
    \mathcal O(\omega^{N_D+1}),
    \nonumber\\
    c^*_n
    &=
    \frac{1}{n!}\sigma_{\mathrm{AC}}^{*(n)}(0)
    -
    \sum_{p=1}^{N^*_{\mathrm{gap}}}
    \frac{R^*_{(0),p}}{(\Gamma^*_{(0),p})^{n+1}} .
\end{align}
Here $N^*_{\mathrm{gap}}$ is the number of gapped poles retained in the transformed EFT. For the fixed effective domain used below, the D3/D5 spectrum displays three regimes as $\tilde{\rho}$ is varied: at low $\tilde{\rho}$ one pole in the original theory maps to one pole in the transformed theory, at intermediate $\tilde{\rho}$ one pole maps to two poles, and at high $\tilde{\rho}$ one pole maps to no pole in the transformed EFT. The corresponding mappings of the coefficients $c_n^*$ were derived in Section~\ref{sec:quasihydro} and Appendix~\ref{app:coefffinite}.

\begin{figure}[h!]
    \centering
    \begin{subfigure}{0.32\textwidth}
        \centering
        \includegraphics[width=\linewidth]{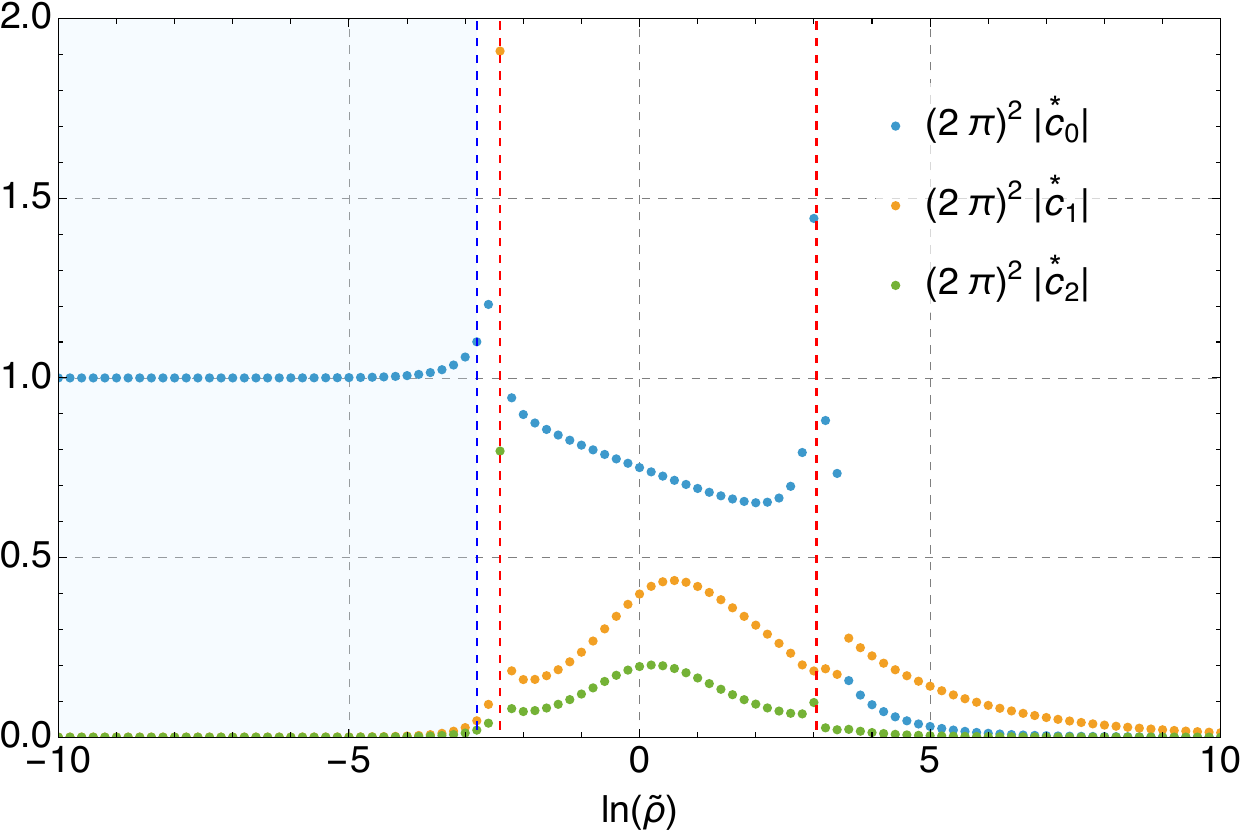}
        \caption{One-to-one regime}
        \label{fig:one_to_one_coefficients}
    \end{subfigure}
    \hfill
    \begin{subfigure}{0.32\textwidth}
        \centering
        \includegraphics[width=\linewidth]{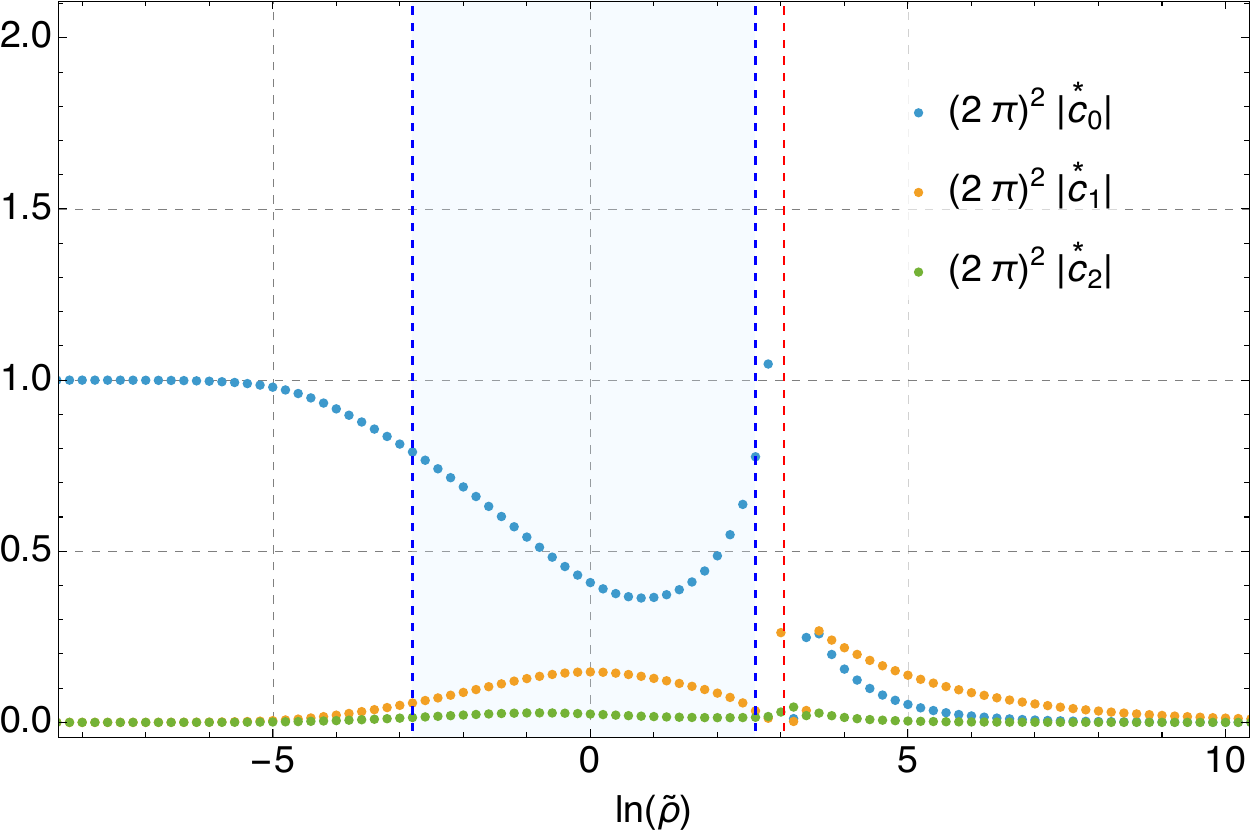}
        \caption{One-to-two regime}
        \label{fig:one_to_two_coefficients}
    \end{subfigure}
    \hfill
    \begin{subfigure}{0.32\textwidth}
        \centering
        \includegraphics[width=\linewidth]{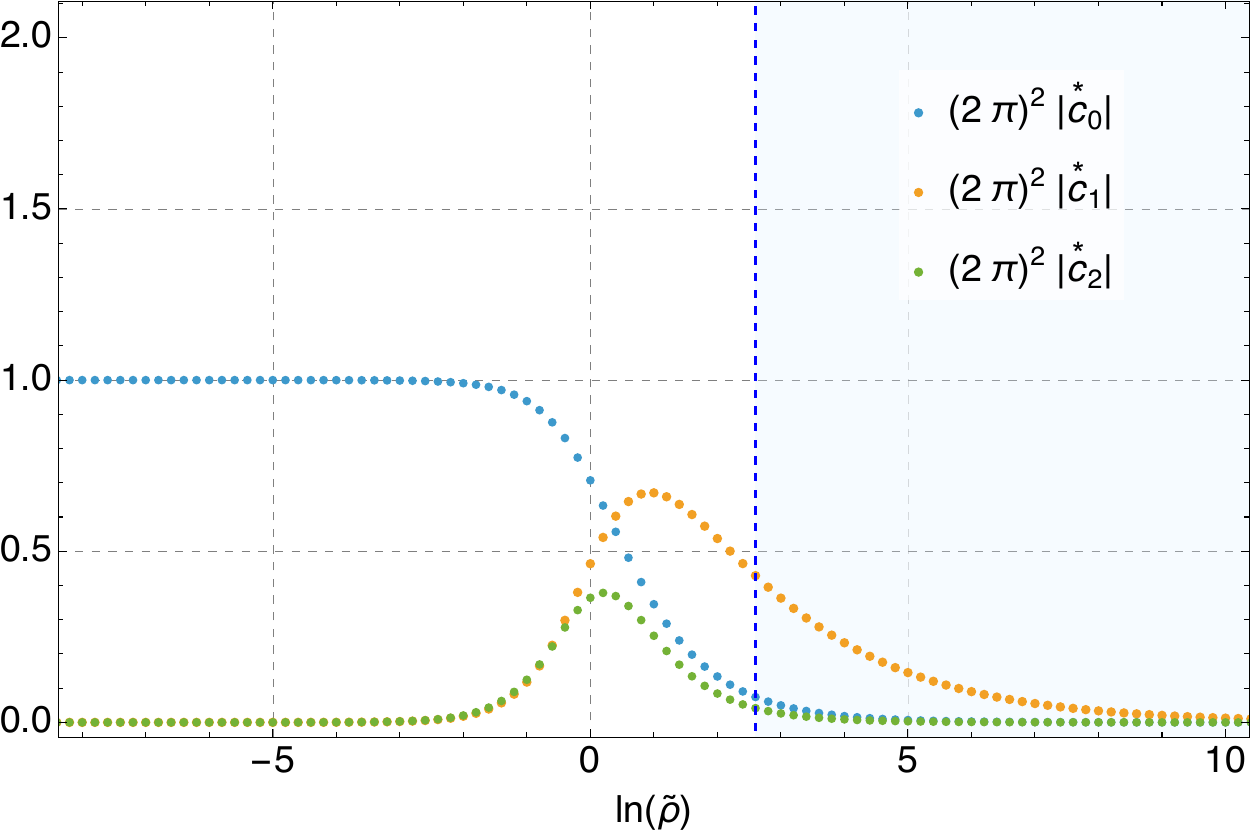}
        \caption{One-to-zero regime}
        \label{fig:one_to_zero_coefficients}
    \end{subfigure}
    \caption{First coefficients $c_n^*$ multiplied by $(2 \pi)^2$ in the expansion of the holomorphic part of the transformed AC conductivity around $\omega=0$ as defined in \ref{eq:coefficients}, plotted as functions of $\tilde{\rho}$. Panel (a) shows the low-$\tilde{\rho}$ regime, where one pole in the original theory maps to one pole in the transformed theory. Panel (b) shows the intermediate-$\tilde{\rho}$ regime, where one pole maps to two transformed poles. Panel (c) shows the high-$\tilde{\rho}$ regime, where one pole maps to no transformed pole inside the chosen effective domain. The blue regions denote the boundaries of the regime of validity for the respecting values of $\tilde{\rho}$. The red vertical lines mark divergences associated with collisions between retained poles and poles outside the truncated domain.}
    \label{fig:coefficients}
\end{figure}

As reviewed above, in a parity-invariant theory the $S$-transformed conductivity is the inverse of the original one, up to the conventional factor of $(2\pi)^2$,
\begin{equation}
\label{eq:DualConductivity}
    \sigma^*_{\mathrm{AC}}(i\omega)
    =
    \frac{1}{(2\pi)^2}
    \frac{1}{\sigma_{\mathrm{AC}}(i\omega)} .
\end{equation}
At large charge density, the leading behaviour of the conductivity in the original theory is Drude-like~\cite{Chen:2017dsy},
\begin{equation}
    \sigma_{\mathrm{Drude}}(i\omega)
    =
    \frac{\chi_{\rho\rho}\tau}{1-i\omega\tau}
    \xrightarrow[\omega\to0]{}
    \chi_{\rho\rho}\tau .
\end{equation}
In the corresponding large-$\tilde{\rho}$ regime of the transformed theory, no gapped pole lies inside the chosen effective domain at $\vec k=\vec0$. The leading low-frequency coefficient is therefore obtained from the one-to-zero map derived in Section~\ref{sec:quasihydro},
\begin{equation}
\label{eq:AClargerho}
    \sigma^*_{\mathrm{AC}}(i\omega)
    \xrightarrow[\omega\to0]{}
    c_0^*
    =
    \frac{\Gamma_{(0),1}}
    {(2\pi)^2\left(c_0\Gamma_{(0),1}+R_{(0),1}\right)} .
\end{equation}
This reproduces the inverse Drude limit implied by~\eqref{eq:DualConductivity}. Indeed, the Drude form corresponds, in the pole-data language, to a single relaxation pole with
\begin{equation}
\label{eq:DrudeIdentification}
    \tau=\frac{1}{\Gamma_{(0),1}},
    \qquad
    \chi_{\rho\rho}=R_{(0),1},
    \qquad
    c_n=0 ,
\end{equation}
exactly the identifications used for the untransformed Drude conductivity in~\cite{Amoretti:2025kem}. Substituting~\eqref{eq:DrudeIdentification} into~\eqref{eq:AClargerho} yields
\begin{equation}
\label{eq:InverseDrude}
    c_0^*
    =
    \frac{1}{(2\pi)^2\,\chi_{\rho\rho}\tau} ,
\end{equation}
which agrees with the direct inversion $\sigma^*_{\mathrm{AC}}(0)=[(2\pi)^2\sigma_{\mathrm{Drude}}(0)]^{-1}$ of~\eqref{eq:DualConductivity}. The overall sign in~\eqref{eq:InverseDrude} is fixed by the convention~\eqref{eq:currentCorrelatorZerok2} relating the retarded current-current correlator to the optical conductivity; the physical DC conductivity of the transformed theory is $c_0^*$, and it matches the numerically computed inverse Drude weight shown in Figure~\ref{fig:coefficients}.

\begin{figure}
    \centering
    \includegraphics[width=0.5\linewidth]{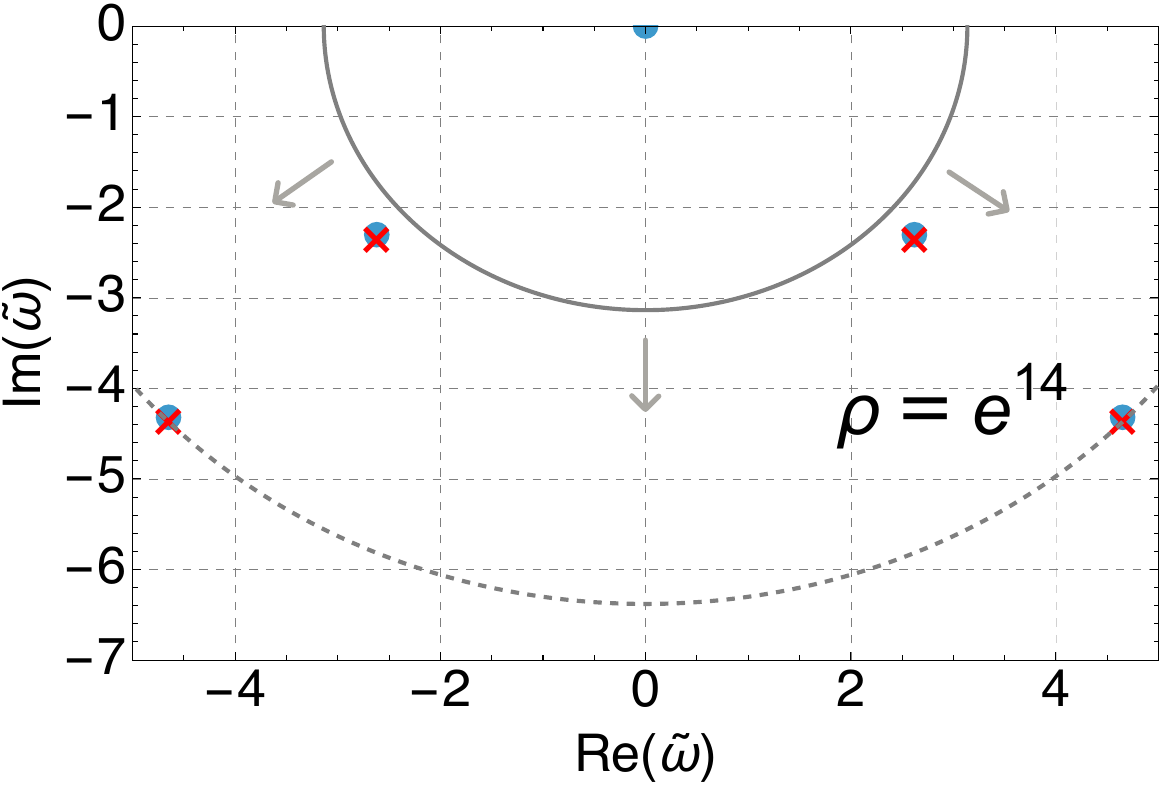}
    \caption{To account for corrections to the inverse Drude form of the transformed conductivity, the effective domain must be enlarged. In the enlarged domain shown here, the original EFT contains three gapped poles, while the transformed EFT contains two. The additional transformed pole data then contribute to the correction terms.
    The positions of the gapped poles of the original theory are shown as blue dots while those of the $S$-transformed theory are shown as red crosses. The dashed grey semi-circle indicates the enlarged frequency domain.   
    }
    \label{fig:threepoles}
\end{figure}

In the original theory, keeping one gapped pole gives access to quasihydrodynamic corrections to the Drude form through the density dependence of the pole residue. In the large-$\tilde{\rho}$ transformed regime considered here, however, no gapped pole is retained inside the chosen domain, so there is no residue contribution within this truncated EFT that can correct the inverse Drude form. To capture such corrections one must enlarge the effective domain. In practice, this means retaining three gapped poles in the original effective description and two gapped poles in the transformed one, as illustrated in Figure~\ref{fig:threepoles}.

We computed the first five coefficients $c_n^*$ in the three density regimes by evaluating the holographic AC conductivity near $\omega=0$. The gapped pole positions were determined from the zeros of the denominator of the holographic response function. We then expanded around each pole to extract its residue, subtracted the retained pole contributions, and finally expanded the remaining holomorphic part in powers of $\tilde\omega$. Figure~\ref{fig:coefficients} shows the resulting coefficients $c_n^*$ as functions of $\tilde{\rho}$. In the one-to-one and one-to-two regimes, divergences appear when retained poles collide with the next poles that were left outside the effective domain. These divergences signal the boundary of validity of the corresponding pole-truncated EFT.

\section{Discussion}
\label{sec:discussion}

We have studied how $SL(2,\mathbb Z)$ transformations act on pole-truncated linear-response effective theories in $2+1$ dimensions. In the purely hydrodynamic sector, the $S$-transformation closes directly on the transport coefficients of the original EFT. Once long-lived non-hydrodynamic modes are retained, the transformed transport coefficients generally require additional infrared data: the positions and, in general, residues of the poles retained in the transformed correlator. The one-to-zero regime is exceptional because no transformed gapped-pole data are required. We derived this structure explicitly for static susceptibilities, the Einstein relation and zero-momentum AC transport.

The alternatively quantised D3/D5 system provides a concrete realisation of these possibilities. Varying $\tilde{\rho}$, equivalently the transformed background magnetic field, changes the pole content inside a fixed effective domain, producing regimes in which an EFT retaining one original gapped pole is associated with transformed EFTs containing one, two or no gapped poles. The agreement between the holographic response and the analytic mappings provides a non-trivial check of the construction, while the large-$\tilde{\rho}$ regime exhibits the expected inverse-Drude behaviour. More generally, the example makes explicit that the mismatch between transforming an EFT and constructing the EFT of the transformed theory is a consequence of finite infrared truncation.

There are several natural directions in which this construction could be
extended. The first is to go beyond the pure $S$-transformation. We have written
the general $SL(2,\mathbb{Z})$ action on currents and sources, and derived the
leading static susceptibility maps for a general transformation. Extending the
finite-frequency analysis to arbitrary $SL(2,\mathbb{Z})$ elements would clarify
how the pole-zero structure is reorganised when the transformation includes both
gauging and Chern-Simons contact terms. This would be particularly relevant in
parity-violating systems, where longitudinal and Hall response are mixed and the
conductivity is naturally matrix-valued.

A second direction is to analyse the full finite-momentum problem. In this paper
we focused on static susceptibilities at non-zero momentum and on AC transport at
zero momentum. At finite $\omega$ and finite $k$, the longitudinal and transverse
sectors contain richer analytic structure, including the interplay between
diffusive poles, gapped relaxation poles, and momentum-dependent residues. The
same logic should apply: the transformed EFT should be reconstructed from the
pole content of the transformed correlators inside the chosen domain. However,
the classification of possible pole rearrangements is expected to be richer than
at $k=0$.

It would also be useful to understand more systematically the dependence on the
choice of effective domain. The number of retained poles is part of the definition
of the EFT, and changing the frequency cutoff changes the data that are kept
explicitly. From this viewpoint, the transformed theory provides a useful
diagnostic of the robustness of a proposed low-energy truncation. If a small
change of domain changes the pole content after duality, the transformed EFT may
require additional degrees of freedom even when the original truncation appears
stable. This suggests a possible criterion for choosing effective domains in
quasihydrodynamic systems: the domain should be large enough to include the pole
data needed for the desired class of duality transformations.

Another natural extension concerns the relation between pole-truncated EFTs and
rational approximations to Green's functions. The effective theory used here
reproduces a Mittag-Leffler expansion of the correlators inside the low-frequency
domain. Since the $S$-transformation maps poles to zeros and zeros to poles, it
acts naturally on the rational structure of the response function. This suggests
that duality may provide a useful organisational principle for constructing
effective descriptions directly at the level of pole-zero data, rather than only
in terms of local transport coefficients.

From the holographic perspective, it would be interesting to repeat the analysis
in models with a richer quasinormal-mode spectrum. The D3/D5 probe brane system
is useful because both quantisations can be treated explicitly and the normal quantisation charge
density provides a clean control parameter. More general holographic systems,
including models with momentum relaxation and incoherent
transport~\cite{Davison:2015bea,Grozdanov:2018fic}, magnetic fields~\cite{Amoretti:2021fch,Amoretti:2021lll,Amoretti_2022},
or spontaneous symmetry breaking and (pseudo-)Goldstone
modes~\cite{Amoretti:2016bxs,Amoretti:2018tzw,Amoretti:2019cef,Amoretti:2019kuf}, should exhibit additional pole
rearrangements under alternative boundary conditions. Such examples could test
whether the one-to-one, one-to-two and one-to-zero regimes found here are part of a
broader pattern.

A further direction is to extend the discussion beyond linear response. The
present analysis is formulated entirely in terms of two-point functions, where
the $SL(2,\mathbb Z)$ action is particularly transparent. Nonlinear
quasihydrodynamic theories containing explicit relaxation modes can be formulated
within the Schwinger-Keldysh framework, where dynamical KMS symmetry relates
nonlinear response to the corresponding stochastic sector~\cite{Amoretti:2026nonlin}.
Beyond the Gaussian level, the noise data are additionally subject to positivity
constraints required for a genuine stochastic probability distribution
~\cite{Amoretti:2026positivenoise}. A nonlinear extension of the present construction would
therefore have to specify not only the transformed pole content of retarded
correlators, but also the corresponding higher-point response and noise data,
together with their KMS and positivity constraints.

Finally, our results emphasise a general conceptual point. Dualities are exact
statements about microscopic theories or complete correlation functions. Effective
theories, by contrast, retain only a finite amount of infrared information. When
the retained data include non-hydrodynamic poles, the operation of truncating to
the infrared does not necessarily commute with the duality transformation. The
duality is exact, but the truncation is not. The transformed effective theory is
therefore obtained not by transforming a finite list of original EFT parameters,
but by transforming the full response function and then rebuilding the EFT from
the pole data that remain inside the effective domain. The value of the present work is that this rebuilding need not be done anew
for every problem: we have packaged it into a closed set of maps. Once the pole
content of the transformed correlator inside the effective domain is identified,
these maps determine the remaining dual EFT data from the original EFT data and the
required transformed pole information. In this precise sense the
$SL(2,\mathbb{Z})$ duality acts on the quasihydrodynamic effective theory once its
non-perturbative pole data are included among the defining infrared data. The pure
$S$-transformation of charge transport is now under full analytic control and
confirmed holographically, and the same pole-data philosophy can be tested further in
the natural next steps, namely general $SL(2,\mathbb{Z})$ elements, finite momentum, and
the nonlinear regime.

\appendix

\section{$SL(2,\mathbb{Z})$ transformation of the generating functional and current transformations} \label{app:sl2z}

{\noindent In this section, we review the effect of the action of $SL(2,\mathbb{Z})$ transformations on the generating functional of $(2+1)$-dimensional QFTs and derive the induced transformations of the corresponding fields and currents stated in Section \ref{section:S-Field}.
}

{We consider a theory with a conserved, gauge-invariant $U(1)$ current $J^\mu$, coupled to a background gauge field $A_\mu$. The generating functional is given by
\begin{equation} \label{eq:GeneratingFunctionalOriginal}
    Z[A] = \int \mathcal{D}\Phi \exp \left( i \, S[\Phi] + i \int d^{(2+1)}x \, J^\mu A_\mu \right) , 
\end{equation}
where $\Phi$ collectively denotes the dynamical fields of the theory and $S[\Phi]$ is the corresponding action. At this stage, $A_\mu$ is treated as a non-dynamical background field.
}

{Following \cite{Witten:2003ya,Leigh:2003ez}, one may construct a new theory by promoting $A_\mu$ to a dynamical field by integrating over it in the path integral. Simultaneously, a new background gauge field, denoted $A_\mu^*$, is introduced and coupled to the conserved current. The generating functional of the transformed theory then becomes
\begin{equation}
    Z[\tilde{J}] = \int \mathcal{D}A \, Z[A] \exp \left( i \int d^{(2+1)}x \, \tilde{J}^\mu A_\mu \right) ,
\end{equation}
where the current $\tilde{J}^\mu$ is defined by
\begin{equation}
    \tilde{J}^\mu = \frac{1}{2\pi} \epsilon^{\mu \nu \rho} \partial_\nu A^*_\rho .
\end{equation}
Upon integrating by parts and neglecting boundary contributions, the generating functional can be written equivalently as 
\begin{eqnarray} \label{eq:GeneratingFunctionalIntermediate}
    Z[A^*] = \int \mathcal{D}A \,\mathcal{D}\Phi \, \exp \left( i S[\Phi] + i S[A] \right), 
\end{eqnarray}
with 
\begin{eqnarray} \label{eq:ActionA}
    S[A] = \int d^{(2+1)} x \, J^\mu A_\mu + \frac{1}{2\pi} \int d^{(2+1)}x \, \epsilon^{\mu \nu \rho} A^*_\mu \partial_\nu A_\rho  \, .
\end{eqnarray}
This expression may be recast in the form
\begin{equation} \label{eq:GeneratingFunctionalNew}
    Z[A^{*}] = \int \mathcal{D}A \; Z[A] \exp \left( i \int d^{(2+1)}x \; J^{\mu *} A^*_{\mu} \right) \; ,       
\end{equation}
where the dual current is defined by
\begin{equation}
J^{\mu *} = \frac{1}{2\pi} \epsilon^{\mu \nu \rho} \partial_\nu A_\rho \, .
\end{equation}
It is convenient to introduce the magnetic current of the original theory, 
\begin{equation} \label{eq:magneticcurrent}
    \mathcal{B}^\mu = \frac{1}{2\pi} \epsilon^{\mu \nu \rho} \partial_\nu A_\rho, 
\end{equation}
in terms of which the dual current satisfies 
\begin{eqnarray} \label{eq:J*Brelation}
    J^{\mu*} = \mathcal{B}^\mu. 
\end{eqnarray}
}

{\noindent Analogously, in the transformed theory we define the dual magnetic current as 
\begin{eqnarray}
    \mathcal{B}^{\mu*} = \frac{1}{2\pi} \epsilon^{\mu \nu \rho} \partial_\nu A^*_\rho. 
\end{eqnarray}
Let us note here that with the convention $\epsilon_{txy} = 1$, we have in component form
\begin{subequations}
    \begin{eqnarray} \label{eq:MagneticCurrent}
        \mathcal{B}^{\mu(*)} &=& \frac{1}{2\pi} \left( -B^{(*)},  \epsilon^{ij} E_j^{(*)} \right).
    \end{eqnarray}
\end{subequations}
To determine the relation of $\mathcal{B}^{\mu *}$ to the original charge current, we vary the $A_\mu$ dependent part of the action \eqref{eq:ActionA} with respect to $A_\mu$. This gives
\begin{eqnarray} \label{eq:JB*relation} 
    \frac{\delta S}{\delta A_\mu} = J^\mu + \mathcal{B}^{*\mu} = 0 \; ,
\end{eqnarray}
where in the last equation we have used the equation of motion of the gauge field, which has been promoted to be dynamical. Combining \eqref{eq:J*Brelation} and \eqref{eq:JB*relation}, we find that the fields and currents of the transformed and original theories are related by 
\begin{equation} \label{eq:S-Transformation}
    \begin{pmatrix}
        J^{\mu *} \\
        \mathcal{B}^{\mu *} 
    \end{pmatrix} = 
    \begin{pmatrix}
        0 & 1 \\
        -1 & 0
    \end{pmatrix}
    \begin{pmatrix}
        J^\mu \\
        \mathcal{B}^\mu  
    \end{pmatrix}.
\end{equation}
}

{In this construction, $A_\mu$ has been promoted to a dynamical gauge field, while $A_\mu^*$ plays the role of an external source. As stated in Section \ref{section:S-Field}, the operation of gauging the original background field and introducing a dual current coupled to a new background field is known as $S$-transformation. The resulting generating functional retains the same functional form as the original expression in \eqref{eq:GeneratingFunctionalOriginal}.
}

{This implementation should be distinguished from electromagnetic self-duality of
a bulk action. In a Maxwell theory in $AdS_4$, the boundary $S$-operation is
closely related to bulk electric-magnetic duality. In the D3/D5 probe brane
system, however, we only use the boundary-condition implementation of the
field-theory $S$-operation. The probe brane action is not assumed to be
invariant under this transformation.}

{In $(2+1)$-dimensions, there is another way to modify the theory by adding a Chern-Simons term for the background gauge field such that the generating functional becomes
\begin{eqnarray}
    Z[A] = \int \mathcal{D}\Phi \exp \left( i S[\Phi] + i \int d^{(2+1)}x \left( J^\mu A_\mu + \frac{n}{4 \pi} \, \epsilon^{\mu \nu \rho} A_\mu \partial_\nu A_\rho \right) \right) \, ,
\end{eqnarray}
with $n$ being integer.
}

{Since the added term depends explicitly on the gauge field $A_\mu$, the current acquires an additional contribution. Taking the functional derivative with respect to $A_\mu$ yields a term proportional to the field strength, motivating the definition of a transformed current,
\begin{eqnarray}
    J^{\mu *} = J^\mu + \frac{n}{2\pi} \, \epsilon^{\mu \nu \rho} \partial_\nu A_\rho \; .
\end{eqnarray}
 We can express this in terms of the hodge dual of the background field strength \eqref{eq:magneticcurrent} as
\begin{eqnarray}
    J^{\mu *} = J^\mu + n \mathcal{B}^\mu \; ,
\end{eqnarray}
This operation of adding a Chern-Simons term term for the background gauge field is called $T$-transformation. We can write the  mixing of the electric current $J^\mu$ and the hodge dual vector as 
\begin{eqnarray}
    \begin{pmatrix}
        J^{\mu *} \\
        \mathcal{B}^{\mu *} 
    \end{pmatrix} = 
    \begin{pmatrix}
        1 & n \\
        0 & 1 
    \end{pmatrix}
    \begin{pmatrix}
        J^\mu \\
        \mathcal{B}^\mu 
    \end{pmatrix} \; .
\end{eqnarray}
}

{The $S$-and $T$-transformations do not commute with each other. Together they generate the $SL(2,\mathbb{Z})$ group. A general $SL(2,\mathbb{Z})$ transformation then acts on the charge and magnetic currents as 
\begin{subequations} \label{eq:SL-Transformationappendix}
    \begin{eqnarray}
        \begin{pmatrix}
            J^{\mu *} \\
            \mathcal{B}^{\mu *} 
        \end{pmatrix} = 
        \begin{pmatrix}
            a & b \\
            c & d
        \end{pmatrix}
        \begin{pmatrix}
            J^\mu \\
            \mathcal{B}^\mu 
        \end{pmatrix} \; ,
    \end{eqnarray} 
    subject to the constraint 
    \begin{eqnarray}
        a d - b c = 1; \quad \quad a,b,c,d \in \mathbb{Z} \, . 
    \end{eqnarray}
\end{subequations}
Written in components, this gives equations \eqref{eq:SL-Field-Transformations} and \eqref{eq:Field-SL-Transformations}. 
}

\section{Additional coefficient mappings} \label{app:mappings}

{In this appendix, we provide higher order mappings of $S$-transformed transport coefficients. Appendix \ref{app:mappingchargesus} and \ref{app:mappingmagsusc} consider the regime of finite wavevector $\vec{k} \neq 0$ and vanishing frequency $\omega = 0$ and provide the mappings for charge and magnetic susceptibilities. Instead, Appendix \ref{app:coefffinite} considers the regime of vanishing wavevector and finite frequency. 
}

\subsection{Charge susceptibility} \label{app:mappingchargesus}

{\begin{equation}
    \begin{split}
        \chi_{\text{EE}}^{*(\text{L})(1)}(0) = &  \frac{-6 \chi _{\text{BB}}(0) \chi _{\text{BB}}'(0) \chi _{\text{BB}}''(0)+6 \chi _{\text{BB}}'(0){}^3+\chi _{\text{BB}}(0){}^2 \chi _{\text{BB}}{}^{(3)}(0)}{24 \pi ^2 \chi _{\text{BB}}(0){}^4} \\
        \chi_{\text{EE}}^{*(\text{L})(2)}(0) = &
        \Bigl( 36 \chi _{\text{BB}}(0) \chi _{\text{BB}}'(0){}^2 \chi _{\text{BB}}''(0)-24 \chi _{\text{BB}}'(0){}^4-8 \chi _{\text{BB}}(0){}^2 \chi _{\text{BB}}{}^{(3)}(0) \chi _{\text{BB}}'(0) \\ 
        & +\chi _{\text{BB}}(0){}^2 \left(\chi _{\text{BB}}(0) \chi _{\text{BB}}{}^{(4)}(0)-6 \chi _{\text{BB}}''(0){}^2\right) \Bigr) \frac{1}{48 \pi ^2 \chi _{\text{BB}}(0){}^5} \\
        \chi_{\text{EE}}^{*(\text{L})(3)}(0) = & \Bigl( \chi _{\text{BB}}(0){}^4 \chi _{\text{BB}}{}^{(5)}(0)-10 \bigl(24 \chi _{\text{BB}}(0) \chi _{\text{BB}}'(0){}^3 \chi _{\text{BB}}''(0) -12 \chi _{\text{BB}}'(0){}^5 \\ 
        & +\chi _{\text{BB}}(0){}^2 \chi _{\text{BB}}'(0) \left(\chi _{\text{BB}}(0) \chi _{\text{BB}}{}^{(4)}(0)-9 \chi _{\text{BB}}''(0){}^2\right) \\
        & -6 \chi _{\text{BB}}(0){}^2 \chi _{\text{BB}}{}^{(3)}(0) \chi _{\text{BB}}'(0){}^2+2 \chi _{\text{BB}}(0){}^3 \chi _{\text{BB}}{}^{(3)}(0) \chi _{\text{BB}}''(0)\bigr) \Bigr) \\
        & \times \frac{1}{80 \pi ^2 \chi _{\text{BB}}(0){}^6} \; . 
    \end{split}
\end{equation}
}

\subsection{Magnetic susceptibility} \label{app:mappingmagsusc}

{
\begin{equation}
    \begin{split}
        \chi_{\text{BB}}^{*(1)}(0) = & \frac{\chi^{(0)}_{\text{EE}}}{(2 \pi) ^2 \chi _{\rho \rho }^2} \; , \\ 
        \chi_{\text{BB}}^{*(2)}(0) = & -\frac{(\chi^{(0)}_{\text{EE}})^2-\chi _{\rho \rho } \chi ^{(L)}_{\text{EE}}(0)}{2 \pi ^2 \chi _{\rho \rho }^3} \; , \\ 
        \chi_{\text{BB}}^{*(3)}(0) = & 3\frac{-2 \chi ^{(0)}_{\text{EE}} \chi ^{(L)}_{\text{EE}}(0) \chi _{\rho \rho }+\chi _{\rho \rho }^2 \chi ^{(L)(1)}_{\text{EE}}(0)+(\chi ^{(0)}_{\text{EE}})^3}{2 \pi ^2 \chi _{\rho \rho }^4} \; , \\
        \chi_{\text{BB}}^{*(4)}(0) = & -\Bigl(6 \bigl(2 \chi ^{(0)}_{\text{EE}} \chi _{\rho \rho }^2 \chi ^{(L)(1)}_{\text{EE}}(0)-3 (\chi ^{(0)}_{\text{EE}})^2 \chi ^{(L)}_{\text{EE}}(0) \chi _{\rho \rho }+ {\chi ^{(L)}_{\text{EE}}(0)}^2 \chi _{\rho \rho }^2+{\chi ^{(0)}_{\text{EE}}}^4 \bigr) \\
        & -3 \chi _{\rho \rho }^3 \chi ^{(L)(2)}_{\text{EE}}(0) \Bigr) \frac{1}{\pi ^2 \chi _{\rho \rho }^5} \\
        \chi_{\text{BB}}^{*(5)}(0) = & - \Bigl( 30 \bigl(-3 (\chi ^{(0)}_{\text{EE}})^2 \chi _{\rho \rho }^2 \chi ^{(L)(1)}_{\text{EE}}(0)+\chi ^{(0)}_{\text{EE}} \chi _{\rho \rho }^2 \bigl( \chi _{\rho \rho } \chi ^{(L)(2)}_{\text{EE}}(0)-3 {\chi ^{(L)}_{\text{EE}}(0)}^2 \bigr) \\ 
        & +4 (\chi ^{(0)}_{\text{EE}})^3 \chi ^{(L)}_{\text{EE}}(0) \chi _{\rho \rho }+2 \chi ^{(L)}_{\text{EE}}(0) \chi _{\rho \rho }^3 \chi ^{(L)(1)}_{\text{EE}}(0)- (\chi ^{(0)}_{\text{EE}})^5 \bigr)-5 \chi ^{(L)(3)}_{\text{EE}}(0) \chi _{\rho \rho }^4 \Bigr) \\
        & \times \frac{1}{\pi ^2 \chi _{\rho \rho }^6}
    \end{split}
\end{equation}
Here, $^{(i)}$, for example in $\chi_{\text{EE}}^{*(\text{L})(i)}(0)$, denotes derivatives with respect to $q=k^2$ and should not be confused with $^{(0)}$ denoting the $k$-independent parts of the susceptibilities. Furthermore $^{i}$ denotes a standard power.}

\subsection{Transport coefficients for $\omega \neq 0$, $\vec{k} = 0$} \label{app:coefffinite}

{We state the $S$-transformed transport coefficients for the case of vanishing wavevector $\vec{k}=\vec{0}$ and finite frequency $\omega \neq 0$. As we argued in Subsection \ref{sec:quasihydro}, we need to cover besides the region where we map one pole to two poles, the regions where we have one pole in the original theory and one pole in the transformed theory, as well as the region where we have one pole in the original theory and no poles in the transformed theory. The case of one-to-two mapping has already been dealt with in Subsection \ref{sec:quasihydro}.
}

{In the "one-to-one" case we find the following results for the coefficients
\begin{eqnarray}
    \bar{\sigma}_{(0)}^{*} = \frac{\Gamma_{(0),1} \Gamma_{(0),1}^{*}}{ (2 \pi ) ^2 \bar{\sigma}_{(0)}}
\end{eqnarray}
and 
\begin{equation}
    \begin{split}
        r_{(0)}^{*}(0) = & \frac{1}{(2 \pi) ^2 \bar{\sigma}_{(0)} ^2} \biggl( -\chi _{\text{EE}}^{(0)} \, \Gamma_{(0),1}^2 \, \Gamma_{(0),1}^{*} + \bar{\sigma}_{(0)} \Gamma_{(0),1}^{*} \left(1- (2 \pi ) ^2 \bar{\sigma}_{(0)} \chi _{\text{EE}}^{(0)*} \right) \\
        & + \Gamma_{(0),1} \left(-r_{(0)}(0) \, \Gamma _{(0),1}^{*}+ \bar{\sigma}_{(0)} \right) \biggr)
    \end{split}
\end{equation}
}

{In the "one-to-zero" case we find the following results for the coefficients
\begin{eqnarray}
    \bar{\sigma}_{(0)}^{*} =  \frac{\Gamma_{(0),1}}{ (2 \pi ) ^2 \bar{\sigma}_{(0)}}
\end{eqnarray}
and 
\begin{equation}
    \begin{split}
        r_{(0)}^{*}(0) = & \frac{1}{(2 \pi) ^2 \bar{\sigma}_{(0)} ^2} \biggl( \bar{\sigma}_{(0)} - (2 \pi)^2 \bar{\sigma}_{(0)}^2 \chi_{\text{EE}}^{(0)*} - \Gamma_{(0),1} \left( \Gamma_{(0),1} \chi_{\text{EE}}^{(0)} + r_{(0)}(0) \right) \biggr)
    \end{split}
\end{equation}
}

{We see again that only in the "one-to-zero" case the coefficients of the $S$-transformed theory are completely determined by just those of the original. This is clear, as in this case we do not need to supply any pole information on the S-dual side. In the "one-to-one" case, this is not true.
}

For the other choice of mapping the coefficients $c_n^*$, we obtain for the one-to-two mapping for the next-to-leading coefficients

\begin{equation}
    \begin{split}
    c^*_{1} =& -\frac{c_1 \Gamma _{(0),1}^2+R_{(0),1}}{(2 \pi) ^2 \left(c_0 \Gamma_{(0),1} +R_{(0),1} \right){}^2}-\frac{R^*_{(0),1}}{\Gamma_{(0),1}^{*2}}-\frac{R^*_{(0),2}}{\Gamma_{(0),2}^{*2}} \\
    c^*_{2} =& - \frac{\left(c_0 c_2-c_1^2\right) \Gamma_{(0),1}^3+c_2 \Gamma_{(0),1}^2 R_{(0),1}-2 c_1 \Gamma_{(0),1} R_{(0),1}+c_0 R_{(0),1}}{4 \pi ^2 \left(c_0 \Gamma_{(0),1}+R_{(0),1}\right){}^3}-\frac{R^*_{(0),1}}{\Gamma_{(0),1}^{*3}}-\frac{R^*_{(0),2}}{\Gamma_{(0),2}^{*3}} \\ 
    c^*_{3} =& - \biggl\{ \left(c_1^3-2 c_0 c_2 c_1+c_0^2 c_3\right) \Gamma_{(0),1}^4+2 \left(c_0 c_3-c_1 c_2\right) \Gamma_{(0),1}^3 R_{(0),1} +R_{(0),1} \left(c_1 R_{(0),1}+c_0^2\right) \\
    & +\Gamma_{(0),1}^2 R_{(0),1} \left(c_3 R_{(0),1}+3 c_1^2-2 c_0 c_2\right)-2 \Gamma_{(0),1} R_{(0),1} \left(c_2 R_{(0),1}+c_0 c_1\right) \biggr\} \\ 
    & \times \frac{1}{4 \pi ^2 \left(c_0 \Gamma_{(0),1}+R_{(0),1}\right){}^4} -\frac{R^*_{(0),1}}{\Gamma_{(0),1}^{*4}}-\frac{R^*_{(0),2}}{\Gamma_{(0),2}^{*4}} \\
    c^*_{4} =& -\biggl\{-\left(\left(c_1^4-3 c_0 c_2 c_1^2+2 c_0^2 c_3 c_1+c_0^2 \left(c_2^2-c_0 c_4\right)\right) \Gamma_{(0),1}^5\right) +R_{(0),1} \Bigl(2 c_1 c_0 R_{(0),1} \\
    & +c_2 R_{(0),1}^2+c_0^3 \Bigr) +\left(3 c_2 c_1^2-4 c_0 c_3 c_1+c_0 \left(3 c_0 c_4-2 c_2^2\right)\right) \Gamma_{(0),1}^4 R_{(0),1} \\
    & -\Gamma_{(0),1}^3 R_{(0),1} \left(\left(c_2^2+2 c_1 c_3-3 c_0 c_4\right) R_{(0),1}+4 c_1^3-6 c_0 c_2 c_1+2 c_0^2 c_3\right) \\
    & +\Gamma_{(0),1}^2 R_{(0),1} \left(c_0 \left(3 c_1^2-4 c_3 R _{(0),1} \right)+R_{(0),1} \left(c_4 R_{(0),1} +6 c_1 c_2\right)-2 c_2 c_0^2\right) \\
    & -\Gamma_{(0),1} R_{(0),1} \left(2 c_3 R_{(0),1}^2+\left(3 c_1^2+c_0 c_2\right) R_{(0),1} +2 c_1 c_0^2\right)\biggr\} \frac{1}{4 \pi ^2 \left(c_0 \Gamma_{(0),1}+R_{(0),1} \right){}^5} \\ 
    & -\frac{R^*_{(0),1}}{\Gamma_{(0),1}^{*5}}-\frac{R^*_{(0),2}}{\Gamma_{(0),2}^{*5}} \\
    c^*_{5} =& -\biggl\{ \Bigl(c_1^5-4 c_0 c_2 c_1^3+3 c_0^2 c_3 c_1^2+c_0^2 \left(3 c_2^2-2 c_0 c_4\right) c_1+c_0^3 \left(c_0 c_5-2 c_2 c_3\right)\Bigr) \Gamma_{(0),1}^6 \\ 
    & + 2 \Bigl(-2 c_2 c_1^3+3 c_0 c_3 c_1^2+3 c_0 \left(c_2^2-c_0 c_4\right) c_1+c_0^2 \left(2 c_0 c_5-3 c_2 c_3\right)\Bigr) \Gamma_{(0),1}^5 R_{(0),1} \\
    & +\Gamma_{(0),1}^4 R_{(0),1} \Bigl(3 c_1^2 \left(c_3 R_{(0),1}-4 c_0 c_2\right)+3 c_1 \left(\left(c_2^2-2 c_0 c_4\right) R_{(0),1} + 2 c_3 c_0^2\right) \\
    & +c_0 \left(3 c_0 \left(2 c_5 R_{(0),1} +c_2^2\right)-6 c_2 c_3 R_{(0),1} -2 c_4 c_0^2\right)+5 c_1^4 \Bigr) \\
    & -2 \Gamma_{(0),1} ^3 R_{(0),1} \Bigl(c_0^2 \left(3 c_4 R_{(0),1}-3 c_1 c_2\right)+c_0 \left(-2 c_5 R_{(0),1}^2-3 \left(c_2^2+2 c_1 c_3\right) R_{(0),1}+2 c_1^3\right) \\
    & +R_{(0),1} \left(\left(c_2 c_3+c_1 c_4\right) R_{(0),1}+6 c_2 c_1^2\right)+c_3 c_0^3 \Bigr) \\
    & +\Gamma_{(0),1}^2 R_{(0),1} \Bigl(c_5 R_{(0),1}^3+3 \left(c_2^2+2 c_1 c_3-2 c_0 c_4\right) R_{(0),1}^2+3 \left(2 c_1^3-c_0^2 c_3\right) R_{(0),1} \\
    & + c_0^2 \left(3 c_1^2-2 c_0 c_2\right)\Bigr) -2 \Gamma_{(0),1} R_{(0),1} \Bigl(3 c_1^2 c_0 R_{(0),1}+R_{(0),1}^2 \left(c_4 R_{(0),1}+3 c_1 c_2\right)+c_1 c_0^3 \Bigr) \\
    & +R_{(0),1} \left(3 c_1 c_0^2 R_{(0),1}+c_3 R_{(0),1}^3+\left(c_1^2+2 c_0 c_2\right) R_{(0),1}^2+c_0^4\right) \biggr\} \frac{1}{4 \pi ^2 \left(c_0 \Gamma_{(0),1}+R_{(0),1}\right){}^6}\\
    & -\frac{R_{(0),1}^*}{\Gamma_{(0),1}^{*6}}-\frac{R_{(0),2}^*}{\Gamma_{(0),2}^{*6}}
    \end{split}
\end{equation}
For the other regimes we just have to omit the last terms, depending on how many poles of the $S$-transformed theory we need to include.

\bibliographystyle{JHEP}
\bibliography{refs}

@article{Brattan:2024dfv,
    author = "Brattan, Daniel K. and Matsumoto, Masataka and Baggioli, Matteo and Amoretti, Andrea",
    title = "{Relaxed hydrodynamic theory of electrically driven non-equilibrium steady states}",
    eprint = "2404.05568",
    archivePrefix = "arXiv",
    primaryClass = "cond-mat.stat-mech",
    doi = "10.1103/PhysRevResearch.6.043097",
    journal = "Phys. Rev. Res.",
    volume = "6",
    pages = "043097",
    month = "11",
    year = "2024"
}

@article{Amoretti:2018tzw,
    author = "Amoretti, Andrea and Are{\'a}n, Daniel and Gout{\'e}raux, Blaise and Musso, Daniele",
    title = "{Universal relaxation in a holographic metallic density wave phase}",
    eprint = "1812.08118",
    archivePrefix = "arXiv",
    primaryClass = "hep-th",
    reportNumber = "CPHT-RR116.122018;IFT-UAM/CSIC-18-130; NORDITA 2018-125;",
    doi = "10.1103/PhysRevLett.123.211602",
    journal = "Phys. Rev. Lett.",
    volume = "123",
    number = "21",
    pages = "211602",
    year = "2019"
}

@article{Heller:2020uuy,
    author = "Heller, Michal P. and Serantes, Alexandre and Spali{\'n}ski, Micha{\l} and Svensson, Viktor and Withers, Benjamin",
    title = "{Hydrodynamic gradient expansion in linear response theory}",
    eprint = "2007.05524",
    archivePrefix = "arXiv",
    primaryClass = "hep-th",
    doi = "10.1103/PhysRevD.104.066002",
    journal = "Phys. Rev. D",
    volume = "104",
    number = "6",
    pages = "066002",
    year = "2021"
}

@article{Davison:2014lua,
    author = "Davison, Richard A. and Gout{\'e}raux, Blaise",
    title = "{Momentum dissipation and effective theories of coherent and incoherent transport}",
    eprint = "1411.1062",
    archivePrefix = "arXiv",
    primaryClass = "hep-th",
    reportNumber = "NORDITA-2014-127, SU-ITP-14-28",
    doi = "10.1007/JHEP01(2015)039",
    journal = "JHEP",
    volume = "01",
    pages = "039",
    year = "2015"
}

@article{Davison:2015bea,
    author = "Davison, Richard A. and Gout{\'e}raux, Blaise",
    title = "{Dissecting holographic conductivities}",
    eprint = "1505.05092",
    archivePrefix = "arXiv",
    primaryClass = "hep-th",
    reportNumber = "SU-ITP-15-12",
    doi = "10.1007/JHEP09(2015)090",
    journal = "JHEP",
    volume = "09",
    pages = "090",
    year = "2015"
}

@article{Donos:2018kkm,
    author = "Donos, Aristomenis and Gauntlett, Jerome P. and Griffin, Tom and Ziogas, Vaios",
    title = "{Incoherent transport for phases that spontaneously break translations}",
    eprint = "1801.09084",
    archivePrefix = "arXiv",
    primaryClass = "hep-th",
    reportNumber = "DCPT-17-29, IMPERIAL-TP-2018-JG-01",
    doi = "10.1007/JHEP04(2018)053",
    journal = "JHEP",
    volume = "04",
    pages = "053",
    year = "2018"
}

@article{Jokela:2013hta,
    author = "Jokela, Niko and Lifschytz, Gilad and Lippert, Matthew",
    title = "{Holographic anyonic superfluidity}",
    eprint = "1307.6336",
    archivePrefix = "arXiv",
    primaryClass = "hep-th",
    doi = "10.1007/JHEP10(2013)014",
    journal = "JHEP",
    volume = "10",
    pages = "014",
    year = "2013"
}

@article{Amoretti:2021lll,
    author = "Amoretti, Andrea and Arean, Daniel and Brattan, Daniel K. and Martinoia, Luca",
    title = "{Hydrodynamic magneto-transport in holographic charge density wave states}",
    eprint = "2107.00519",
    archivePrefix = "arXiv",
    primaryClass = "hep-th",
    doi = "10.1007/JHEP11(2021)011",
    journal = "JHEP",
    volume = "11",
    pages = "011",
    year = "2021"
}

@article{Amoretti:2021fch,
    author = "Amoretti, Andrea and Arean, Daniel and Brattan, Daniel K. and Magnoli, Nicodemo",
    title = "{Hydrodynamic magneto-transport in charge density wave states}",
    eprint = "2101.05343",
    archivePrefix = "arXiv",
    primaryClass = "hep-th",
    reportNumber = "IFT-UAM/CSIC-21-1",
    doi = "10.1007/JHEP05(2021)027",
    journal = "JHEP",
    volume = "05",
    pages = "027",
    year = "2021"
}

@article{Amoretti:2016bxs,
    author = "Amoretti, Andrea and Are{\'a}n, Daniel and Argurio, Riccardo and Musso, Daniele and Pando Zayas, Leopoldo A.",
    title = "{A holographic perspective on phonons and pseudo-phonons}",
    eprint = "1611.09344",
    archivePrefix = "arXiv",
    primaryClass = "hep-th",
    doi = "10.1007/JHEP05(2017)051",
    journal = "JHEP",
    volume = "05",
    pages = "051",
    year = "2017"
}

@article{Amoretti:2019cef,
    author = "Amoretti, Andrea and Are{\'a}n, Daniel and Gout{\'e}raux, Blaise and Musso, Daniele",
    title = "{Diffusion and universal relaxation of holographic phonons}",
    eprint = "1904.11445",
    archivePrefix = "arXiv",
    primaryClass = "hep-th",
    reportNumber = "CPHT-RR018.042019;IFT-UAM/CSIC-19-55, CPHT-RR018.042019",
    doi = "10.1007/JHEP10(2019)068",
    journal = "JHEP",
    volume = "10",
    pages = "068",
    year = "2019"
}

@article{Amoretti:2025kem,
    author = "Amoretti, Andrea and Brattan, Daniel K. and Rongen, Jonas",
    title = "{Linear response beyond hydrodynamic poles}",
    eprint = "2512.19694",
    archivePrefix = "arXiv",
    primaryClass = "hep-th",
    doi = "10.1007/JHEP07(2026)092",
    journal = "JHEP",
    volume = "07",
    pages = "092",
    year = "2026"
}

@article{Brattan:2013wya,
    author = "Brattan, Daniel K. and Lifschytz, Gilad",
    title = "{Holographic plasma and anyonic fluids}",
    eprint = "1310.2610",
    archivePrefix = "arXiv",
    primaryClass = "hep-th",
    doi = "10.1007/JHEP02(2014)090",
    journal = "JHEP",
    volume = "02",
    pages = "090",
    year = "2014"
}

@article{Brattan:2014moa,
    author = "Brattan, Daniel K.",
    title = "{A strongly coupled anyon material}",
    eprint = "1412.1489",
    archivePrefix = "arXiv",
    primaryClass = "hep-th",
    doi = "10.1007/JHEP11(2015)214",
    journal = "JHEP",
    volume = "11",
    pages = "214",
    year = "2015"
}

@article{Jokela:2014wsa,
    author = "Jokela, Niko and Lifschytz, Gilad and Lippert, Matthew",
    title = "{Flowing holographic anyonic superfluid}",
    eprint = "1407.3794",
    archivePrefix = "arXiv",
    primaryClass = "hep-th",
    reportNumber = "HIP-2014-13-TH",
    doi = "10.1007/JHEP10(2014)021",
    journal = "JHEP",
    volume = "10",
    pages = "021",
    year = "2014"
}

@article{Itsios:2016ffv,
    author = "Itsios, Georgios and Jokela, Niko and Ramallo, Alfonso V.",
    title = "{Collective excitations of massive flavor branes}",
    eprint = "1602.06106",
    archivePrefix = "arXiv",
    primaryClass = "hep-th",
    reportNumber = "HIP-2016-04-TH",
    doi = "10.1016/j.nuclphysb.2016.06.008",
    journal = "Nucl. Phys. B",
    volume = "909",
    pages = "677--724",
    year = "2016"
}

@article{Jokela:2017fwa,
    author = "Jokela, Niko and Lifschytz, Gilad and Lippert, Matthew",
    title = "{Striped anyonic fluids}",
    eprint = "1706.05006",
    archivePrefix = "arXiv",
    primaryClass = "hep-th",
    reportNumber = "HIP-2017-11-TH",
    doi = "10.1103/PhysRevD.96.046016",
    journal = "Phys. Rev. D",
    volume = "96",
    number = "4",
    pages = "046016",
    year = "2017"
}

@article{Withers_2018,
   title={Short-lived modes from hydrodynamic dispersion relations},
   volume={2018},
   ISSN={1029-8479},
   url={http://dx.doi.org/10.1007/JHEP06(2018)059},
   DOI={10.1007/jhep06(2018)059},
   number={6},
   journal={Journal of High Energy Physics},
   publisher={Springer Science and Business Media LLC},
   author={Withers, Benjamin},
   year={2018},
   month=jun }

@article{Grozdanov:2018fic,
    author = "Grozdanov, Sa{\v{s}}o and Lucas, Andrew and Poovuttikul, Napat",
    title = "{Holography and hydrodynamics with weakly broken symmetries}",
    eprint = "1810.10016",
    archivePrefix = "arXiv",
    primaryClass = "hep-th",
    reportNumber = "MIT-CTP/5075",
    doi = "10.1103/PhysRevD.99.086012",
    journal = "Phys. Rev. D",
    volume = "99",
    number = "8",
    pages = "086012",
    year = "2019"
}

@article{Amoretti_2022,
   title={On the hydrodynamics of (2 + 1)-dimensional strongly coupled relativistic theories in an external magnetic field},
   volume={37},
   ISSN={1793-6632},
   url={http://dx.doi.org/10.1142/S0217732322300105},
   DOI={10.1142/s0217732322300105},
   number={21},
   journal={Modern Physics Letters A},
   publisher={World Scientific Pub Co Pte Ltd},
   author={Amoretti, Andrea and Brattan, Daniel K.},
   year={2022},
   month=jul }

@inproceedings{Witten:2003ya,
    author = "Witten, Edward",
    title = "{SL(2,Z) action on three-dimensional conformal field theories with Abelian symmetry}",
    booktitle = "{From Fields to Strings: Circumnavigating Theoretical Physics: A Conference in Tribute to Ian Kogan}",
    eprint = "hep-th/0307041",
    archivePrefix = "arXiv",
    pages = "1173--1200",
    month = "7",
    year = "2003"
}

@article{Leigh:2003ez,
    author = "Leigh, Robert G. and Petkou, Anastasios C.",
    title = "{SL(2,Z) action on three-dimensional CFTs and holography}",
    eprint = "hep-th/0309177",
    archivePrefix = "arXiv",
    reportNumber = "CERN-TH-2003-215, ILL-TH-03-08",
    doi = "10.1088/1126-6708/2003/12/020",
    journal = "JHEP",
    volume = "12",
    pages = "020",
    year = "2003"
}

@article{Chen:2017dsy,
    author = "Chen, Chi-Fang and Lucas, Andrew",
    title = "{Origin of the Drude peak and of zero sound in probe brane holography}",
    eprint = "1709.01520",
    archivePrefix = "arXiv",
    primaryClass = "hep-th",
    doi = "10.1016/j.physletb.2017.10.023",
    journal = "Phys. Lett. B",
    volume = "774",
    pages = "569--574",
    year = "2017"
}

@article{Dubovsky:2011sj,
    author = "Dubovsky, Sergei and Hui, Lam and Nicolis, Alberto and Son, Dam Thanh",
    title = "{Effective field theory for hydrodynamics: thermodynamics, and the derivative expansion}",
    eprint = "1107.0731",
    archivePrefix = "arXiv",
    primaryClass = "hep-th",
    doi = "10.1103/PhysRevD.85.085029",
    journal = "Phys. Rev. D",
    volume = "85",
    pages = "085029",
    year = "2012"
}

@article{Crossley:2015evo,
    author = "Crossley, Michael and Glorioso, Paolo and Liu, Hong",
    title = "{Effective field theory of dissipative fluids}",
    eprint = "1511.03646",
    archivePrefix = "arXiv",
    primaryClass = "hep-th",
    doi = "10.1007/JHEP09(2017)095",
    journal = "JHEP",
    volume = "09",
    pages = "095",
    year = "2017"
}

@article{Glorioso:2016gsa,
    author = "Glorioso, Paolo and Crossley, Michael and Liu, Hong",
    title = "{Effective field theory of dissipative fluids (II): classical limit, dynamical KMS symmetry and entropy current}",
    eprint = "1701.07817",
    archivePrefix = "arXiv",
    primaryClass = "hep-th",
    doi = "10.1007/JHEP09(2017)096",
    journal = "JHEP",
    volume = "09",
    pages = "096",
    year = "2017"
}

@article{Kovtun:2014hpa,
    author = "Kovtun, Pavel and Moore, Guy D. and Romatschke, Paul",
    title = "{Towards an effective action for relativistic dissipative hydrodynamics}",
    eprint = "1405.3967",
    archivePrefix = "arXiv",
    primaryClass = "hep-th",
    doi = "10.1007/JHEP07(2014)123",
    journal = "JHEP",
    volume = "07",
    pages = "123",
    year = "2014"
}

@article{Baggioli:2023mid,
    author = "Baggioli, Matteo and Bu, Yanyan and Ziogas, Vasilis",
    title = "{U(1) quasi-hydrodynamics: Schwinger-Keldysh effective field theory and holography}",
    eprint = "2304.14173",
    archivePrefix = "arXiv",
    primaryClass = "hep-th",
    doi = "10.1007/JHEP09(2023)019",
    journal = "JHEP",
    volume = "09",
    pages = "019",
    year = "2023"
}

@article{Liu:2024tqe,
    author = "Liu, Yan and Sun, Ya-Wen and Wu, Xin-Meng",
    title = "{Holographic Schwinger-Keldysh effective field theories including a non-hydrodynamic mode}",
    eprint = "2411.16306",
    archivePrefix = "arXiv",
    primaryClass = "hep-th",
    doi = "10.1016/j.physc.2025.1354701",
    journal = "Physica C",
    volume = "632",
    pages = "1354701",
    year = "2025"
}

@article{Grozdanov:2019kge,
    author = "Grozdanov, Sa{\v{s}}o and Kovtun, Pavel K. and Starinets, Andrei O. and Tadi{\'c}, Petar",
    title = "{Convergence of the Gradient Expansion in Hydrodynamics}",
    eprint = "1904.01018",
    archivePrefix = "arXiv",
    primaryClass = "hep-th",
    doi = "10.1103/PhysRevLett.122.251601",
    journal = "Phys. Rev. Lett.",
    volume = "122",
    number = "25",
    pages = "251601",
    year = "2019"
}

@article{Grozdanov:2019uhi,
    author = "Grozdanov, Sa{\v{s}}o and Kovtun, Pavel K. and Starinets, Andrei O. and Tadi{\'c}, Petar",
    title = "{The complex life of hydrodynamic modes}",
    eprint = "1904.12862",
    archivePrefix = "arXiv",
    primaryClass = "hep-th",
    doi = "10.1007/JHEP11(2019)097",
    journal = "JHEP",
    volume = "11",
    pages = "097",
    year = "2019"
}

@article{Turner:2019wnh,
    author = "Turner, Carl",
    title = "{Dualities in 2+1 Dimensions}",
    eprint = "1905.12656",
    archivePrefix = "arXiv",
    primaryClass = "hep-th",
    doi = "10.22323/1.349.0001",
    journal = "PoS",
    volume = "Modave2018",
    pages = "001",
    year = "2019"
}

@article{Metlitski:2015eka,
    author = "Metlitski, Max A. and Vishwanath, Ashvin",
    title = "{Particle-vortex duality of two-dimensional Dirac fermion from electric-magnetic duality of three-dimensional topological insulators}",
    eprint = "1505.05142",
    archivePrefix = "arXiv",
    primaryClass = "cond-mat.str-el",
    doi = "10.1103/PhysRevB.93.245151",
    journal = "Phys. Rev. B",
    volume = "93",
    number = "24",
    pages = "245151",
    year = "2016"
}

@article{Herzog:2007ij,
    author = "Herzog, Christopher P. and Kovtun, Pavel and Sachdev, Subir and Son, Dam Thanh",
    title = "{Quantum critical transport, duality, and M-theory}",
    eprint = "hep-th/0701036",
    archivePrefix = "arXiv",
    doi = "10.1103/PhysRevD.75.085020",
    journal = "Phys. Rev. D",
    volume = "75",
    pages = "085020",
    year = "2007"
}

@book{Sachdev:2011wg,
    author = "Sachdev, Subir",
    title = "{Quantum Phase Transitions}",
    publisher = "Cambridge University Press",
    edition = "2",
    year = "2011"
}

@article{Hartman:2008dq,
    author = "Hartman, Thomas and Rastelli, Leonardo",
    title = "{Double-trace deformations, mixed boundary conditions and functional determinants in AdS/CFT}",
    eprint = "0801.2785",
    archivePrefix = "arXiv",
    primaryClass = "hep-th",
    doi = "10.1088/1126-6708/2008/01/019",
    journal = "JHEP",
    volume = "01",
    pages = "019",
    year = "2008"
}

@article{Amoretti:2019kuf,
    author = "Amoretti, Andrea and Are\'an, Daniel and Gout\'eraux, Blaise and Musso, Daniele",
    title = "{Gapless and gapped holographic phonons}",
    eprint = "1910.11330",
    archivePrefix = "arXiv",
    primaryClass = "hep-th",
    doi = "10.1007/JHEP01(2020)058",
    journal = "JHEP",
    volume = "01",
    pages = "058",
    year = "2020"
}

@article{Amoretti:2026positivenoise,
    author = "Amoretti, Andrea and Brattan, Daniel K.",
    title = "{Staying positive: bounds for non-Gaussian noise in Schwinger-Keldysh effective field theory}",
    eprint = "2609.12050",
    archivePrefix = "arXiv",
    primaryClass = "cond-mat.stat-mech",
    month = "9",
    year = "2026"
}

@article{Amoretti:2026nonlin,
    author = "Amoretti, Andrea and Anselmi, Matteo and Brattan, Daniel K.",
    title = "{The price of locality: Maxwell-Cattaneo charge transport in Schwinger-Keldysh effective field theory}",
    eprint = "2609.13378",
    archivePrefix = "arXiv",
    primaryClass = "cond-mat.stat-mech",
    month = "9",
    year = "2026"
}

@article{Amoretti:2026branch,
    author = "Amoretti, Andrea and Anselmi, Matteo and Brattan, Daniel K.",
    title = "{Schwinger-Keldysh effective actions for non-hydrodynamic poles and branch cuts}",
    eprint = "2609.16164",
    archivePrefix = "arXiv",
    primaryClass = "hep-th",
    month = "9",
    year = "2026"
}

\end{document}